\documentclass[twocolumn]{aastex701}
\usepackage{amsmath}
\usepackage{multirow}
\begin{document}

\title{On-Orbit Calibration of Danuri/PolCam. II. Image Artifact Correction}

\correspondingauthor{Sungsoo S. Kim}

\author[orcid=0000-0002-2703-7810, gname=Kilho, sname=Baek]{Kilho Baek}
% \altaffiliation{School of Space Research}
\affiliation{School of Space Research, Kyung Hee University, Yongin-si, Gyeonggi-do 17104, Republic of Korea}
\email{kilho.baek@khu.ac.kr}

\author[orcid=0000-0002-5570-2160, gname=Sungsoo, sname=Kim]{Sungsoo S. Kim} 
% \altaffiliation{School of Space Research}
\affiliation{Humanitas College, Kyung Hee University, Yongin-si, Gyeonggi-do 17104, Republic of Korea}
\affiliation{Department of Astronomy and Space Science, Kyung Hee University, Yongin-si, Gyeonggi-do 17104, Republic of Korea}
\email[show]{sungsoo.kim@khu.ac.kr}

\author[orcid=0000-0002-5434-5181, gname=Minsup, sname=Jeong]{Minsup Jeong}
\affiliation{Korea Astronomy and Space Science Institute, Daejeon 34055, Republic of Korea}
\email{msjeong@kasi.re.kr}

\author[orcid=0000-0001-6060-5851, gname=Young-Jun, sname=Choi]{Young-Jun Choi}
\affiliation{Korea Astronomy and Space Science Institute, Daejeon 34055, Republic of Korea}
\email{yjchoi@kasi.re.kr}

\begin{abstract}
Danuri, South Korea's first lunar orbiter, was launched on 2022 August~5, and has completed its two-year nominal mission phase. The wide-angle Polarimetric Camera (PolCam) onboard Danuri is the first instrument to conduct global polarimetric observations from lunar orbit. This paper presents the image artifact correction pipeline for PolCam's on-orbit data. It encompasses the primary steps of the radiometric calibration, consisting of dark current removal, smear correction, and flat-fielding, except for inter-channel relative response and the final radiance calculation. Notably, PolCam's raw data exhibit severe smear artifacts induced by the frame-transfer CCD architecture, which degrade both radiometric fidelity and the accuracy of polarimetric measurements. These smear artifacts have been mitigated through a rigorous correction algorithm, restoring data quality to a level sufficient for scientific analysis and facilitating the precise derivation of the degree of linear polarization (DoLP). These steps serve as an essential prerequisite for conducting subsequent absolute radiance and photometric calibrations. Finally, we present representative examples of polarimetric measurements to validate the calibration performance. Although the current calibration focuses on restoring data quality for qualitative scientific analysis, these results demonstrate the expected inverse linear relationship between intensity and polarization.
\end{abstract}


\section{Introduction}\label{sec:intro}

Danuri, also known as the Korea Pathfinder Lunar Orbiter (KPLO), was launched on 2022 August~5, and entered its target 100~km orbit on 2022 December~27 \citep{Song2023}. Following an 1-month commissioning phase, the nominal mission phase commenced on 2023 February~4 \citep{Jeon2024}. The wide-angle Polarimetric Camera (PolCam), one of the payloads onboard Danuri, underwent an additional $\sim$$2$~months of performance verification before commencing full-scale scientific observations on 2023 March~26. After the nominal mission phase concluded on 2025 February~18, the first extended mission phase was initiated on 2025 February~19. During this extended phase, observations were conducted from a lower 60~km orbit for $\sim$$7$~months before the spacecraft transitioned into a frozen orbit on 2025 September~24.

\defcitealias{Baek2025}{Paper~I}
The primary objective of PolCam was to conduct the world’s first polarimetric survey of the Moon, producing high-resolution (512~PPD $\sim$ 59~$\mathrm{m\,pixel^{-1}}$) polarization maps of the entire lunar surface within latitudes of $\pm70^{\circ}$ \citep{Sim2020}. Throughout its $\sim$$2$-year nominal mission, PolCam acquired at least five observations for the entire region across a diverse range of phase angles. To generate a reliable global polarization map from this dataset, precise geometric and radiometric calibrations are essential. Geometric calibration was achieved by co-registering with other orbital data, compensating for the absence of pre-launch camera model measurements \citep[][hereafter \citetalias{Baek2025}]{Baek2025}.

Radiometric calibration entails the conversion of raw digital number (DN) into radiance, typically involving dark current removal, flat-fielding, and total optical efficiency corrections for both optical filters and the detector. For instruments utilizing frame-transfer CCDs like PolCam, smear correction is a mandatory step. Frame-transfer CCDs have been employed in several space missions, including MESSENGER's Mercury Dual Imaging System \citep[MDIS;][]{Denevi2018, Hawkins2007, Murchie1999}, Kaguya's Multiband Imager \citep[MI;][]{Kodama2010}, Hayabusa's Asteroid Multi-band Imaging Camera \citep[AMICA;][]{Ishiguro2010}, and the CHaracterising ExOPlanet Satellite \citep[CHEOPS;][]{Hoyer2020}. These preceding missions established their smear correction protocols based on pre-launch ground experiments that characterized charge smearing effects.

\defcitealias{Iglesias2015}{I15}
Charge smearing is a pervasive artifact in shutterless frame-transfer CCDs that necessitates numerical correction to achieve high-precision imaging. While early algorithms established the fundamental baseline for smear removal \citep{Powell1999, Ruyten1999}, they relied on the simplifying assumption that sensor illumination remains constant during both the exposure and readout phases \citep{Knox2007, Feng2013}. To address the limitations of this static assumption, \citet[][hereafter \citetalias{Iglesias2015}]{Iglesias2015} reformulated the de-smearing model to accommodate significant intensity variations between consecutive frames. This dynamic approach has proven highly effective in applications such as fast polarimetric imaging, where intensity levels fluctuate rapidly between successive exposures.

The de-smearing model proposed by \citetalias{Iglesias2015} is well-suited for PolCam, as it incorporates a model for intensity variations and was demonstrated for polarimetry using the Fast Solar Polarimeter \citep[FSP;][]{Feller2014, Iglesias2016}. Although their study mathematically derived models for both non-periodic and periodic illumination, its practical application remained limited to the periodic case because the FSP observes temporal intensity variations within a fixed field of view. In contrast, PolCam continuously scans distinct surface regions due to its orbital motion, inherently generating a non-periodic illumination profile. The correction process for PolCam is more challenging, since it records only 6~lines per frame rather than the full CCD array. This study presents the first experimental validation of the de-smearing algorithm under non-periodic illumination, a scenario formulated by \citetalias{Iglesias2015} but hitherto unverified scenario.

Unfortunately, the pre-launch ground experiments for PolCam proved inadequate for radiometric calibration. Although fundamental measurements for dark current and flat-field were acquired \citep{Jeong2023}, they were inconsistent with the actual on-orbit performance. No laboratory experiment was performed to measure the optical efficiency using the combination of color filters and polarizers used in the flight model. As the charge-smearing effect was not anticipated prior to launch, no ground experimental data exist to characterize this artifact.

Various operational malfunctions were encountered during the early nominal mission phase: certain modes yielded anomalous data comprising only binary values (0 and 1), the binning mode exhibited premature saturation at digital numbers below the full dynamic range, and normal operations were restricted to exposure times ($\le$$5.644~\mathrm{ms}$) much shorter than the optimal exposure time ($8.134~\mathrm{ms}$). To avoid these malfunctions, we identified viable operating conditions and performed observations. Given these deviations from planned operations, we are cautious about applying the pre-launch laboratory test data to the on-orbit calibration. We face the challenge of establishing a radiometric calibration pipeline relying solely on on-orbit observation data.

In the present study, we establish the image artifact correction method required to derive the degree of linear polarization from PolCam observations. It encompasses the primary steps of the radiometric calibration except for inter-channel relative response and final radiance calculation. A notable contribution of this work is the effective mitigation of severe smear artifacts inherent to the frame-transfer CCD architecture. Section~\ref{sec:polcam} describes the color and polarization filter combinations, along with the specific CCD characteristics of PolCam. Section~\ref{sec:inscal} details the image artifact correction process, including dark current removal, flat-fielding, and smear correction. Section~\ref{sec:dolp} demonstrates the polarimetric measurements derived from the the data corrected for geometric and instrumental effects. Section~\ref{sec:condis} provides the summary and concluding discussions.

\section{PolCam}\label{sec:polcam}

Achieving precise calibration requires a thorough understanding of the instrument. The optical system of PolCam consists of a six-lens compound objective, filters, and a Charge-Coupled Device (CCD). While the optical geometry of the lens system was characterized in \citetalias{Baek2025}, this section focuses on the filters and CCD sensor characteristics relevant to image artifact correction.

% F1: CCD & Filters layout -----------------------------------
\begin{figure}[t]
\centering
\includegraphics[width=0.47\textwidth]{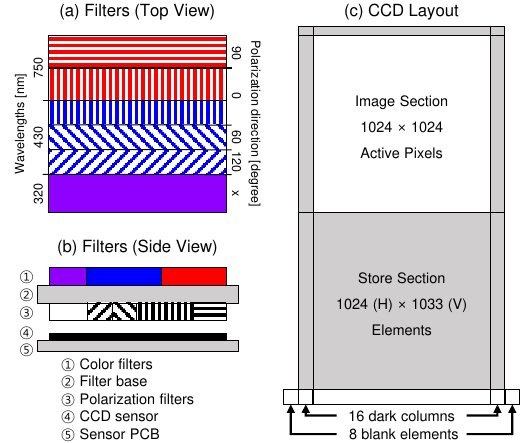}
\caption{Configuration of the PolCam bandpass filters and polarizers shown in (a) top view and (b) side view, alongside (c) a simplified schematic of the CCD structure. In Panel (c), gray shaded regions indicate areas where incident light is physically masked. Note that the 430~nm band is depicted in blue to accurately reflect its spectral color, correcting the green color coding used in previous studies (\citetalias{Baek2025}; \citealt{Jeong2023}).\label{fig:ccd}}
% \vspace*{5mm}
\end{figure}
% ------------------------------------------------------------

\subsection{Filters}\label{sec:filters}

PolCam employs three bandpass filters (320, 430, and 750~nm) and four polarization filters ($0^{\circ}$, $60^{\circ}$, $90^{\circ}$, and $120^{\circ}$). These are configured into a total of six channels: the 320~nm band operates without a polarization filter, while the 430~nm band is paired with polarizers at $0^\circ$, $60^\circ$, and $120^\circ$, and the 750~nm band with $0^\circ$ and $90^\circ$ (Figure~\ref{fig:ccd}(a)). The color filters are positioned in front of the polarization filters (i.e., closer to the lens) (Figure~\ref{fig:ccd}(b)). The channels are arranged on the CCD from the bottom (nearest to the output register adjacent to blank elements) to the top as follows: 320~nm (Ch.~1), 430~nm with $120^\circ$ (Ch.~2), 430~nm with $60^\circ$ (Ch.~3), 430~nm with $0^\circ$ (Ch.~4), 750~nm with $0^\circ$ (Ch.~5), and 750~nm with $90^\circ$ (Ch.~6) (\citetalias{Baek2025}; \citealt{Jeong2023}). This sequence corresponds to the layout of the image section shown in Figure~\ref{fig:ccd}(c).

Conventional polarimetric measurements employ four polarizer angles ($0^{\circ}$, $45^{\circ}$, $90^{\circ}$, and $135^{\circ}$), necessitating 8~channels for two spectral bands. To minimize data volume, PolCam adopted an optimized set of polarizer angles: $0^{\circ}$, $60^{\circ}$, and $120^{\circ}$ for the 430~nm band, and $0^{\circ}$ and $90^{\circ}$ for the 750~nm band. The degree and angle of linear polarization can be derived using the $0^{\circ}$, $60^{\circ}$, and $120^{\circ}$ observations (Equations~(\ref{eq:I430})--(\ref{eq:P430})). If the angle of linear polarization is known, the degree of linear polarization can be retrieved using the $0^{\circ}$ and $90^{\circ}$ measurements (Equations~(\ref{eq:A430}) and~(\ref{eq:P750})). The detailed derivation based on the generalized Mueller matrix is described in Appendix~\ref{sec:calcdop}.

\subsection{CCD}\label{sec:ccd}

PolCam is equipped with the \texttt{AIMO} (Advanced Inverted Mode Operation) version of the \texttt{Teledyne e2v CCD47-20} sensor with a $1024\times1024$ pixel format and a pixel size of $13\times13~\mathrm{{\mu}m}$ (\citetalias{Baek2025}; \citealt{Jeong2023}). The key physical characteristics and nominal performance parameters of the sensor are summarized in Table~\ref{tab:ccd}. This back-illuminated sensor features a frame-transfer split architecture, comprising an exposed image section and a masked storage section, as shown in Figure~\ref{fig:ccd}(c). Accumulated charges are rapidly shifted to the storage section for readout, enabling the simultaneous exposure of the next frame. While this shutterless design offers high frame rates and a continuous duty cycle, it is susceptible to charge smearing caused by incident light during the vertical charge transfer.

% T1: CCD ----------------------------------------------------
\begin{deluxetable}{lcc}
\centering\setlength{\tabcolsep}{2pt}
\tablecaption{Specifications of the Teledyne e2v CCD47-20 (AIMO) sensor.\label{tab:ccd}}
\tablehead{
 & Typical Value & Unit
}
\startdata
Active image format & $1024 \times 1024$ & pixels \\
Pixel Size & $13 \times 13$ & $\mathrm{\mu m}$ \\
Image Area & $13.3 \times 13.3$ & $\mathrm{mm}$ \\
Peak charge storage & 100 & $\mathrm{ke^{-}\,pixel^{-1}}$ \\
Peak output voltage & 450 & $\mathrm{mV}$ \\
Dark signal at $293~\mathrm{K}$ & 250 & $\mathrm{e^{-}\,pixel^{-1}\,s^{-1}}$ \\
Dynamic range & 50,000:1 & --- \\
Charge transfer efficiency & & \\
\quad Parallel & $99.9999$ & \% \\
\quad Serial & $99.9993$ & \% \\
Output amplifier responsivity & 4.5 & $4.5~\mathrm{\mu V/e^{-}}$ \\
Readout noise at $253~\mathrm{K}$ & 2.0 & $\mathrm{rms\ e^{-}\,pixel^{-1}}$ \\
Maximum readout frequency & 5.0 & $\mathrm{MHz}$ \\
Dark signal non-uniformity & 60 & $\mathrm{e^{-}\,pixel^{-1}\,s^{-1}}$
\enddata
\tablecomments{The manufacturer's official specifications.}
\vspace*{-3mm}
\end{deluxetable}
% ------------------------------------------------------------

During signal digitization, the system gain is set by the equation $\mathrm{Gain} = 6 / [1 + 5 \times (63 - G) / 63]$, where $G$ is a configurable parameter, ranging from 0 to 63, set via mission commands. The sensor has an intrinsic analog dynamic range of 50,000:1. Standard operations employ a fixed gain of $1~\mathrm{e^{-}/ADU}$ ($G = 0$). This setting is chosen to provide the sensitivity needed to capture weak polarimetric signals from the lunar surface. The effective system-level dynamic range is limited to $2^{14}$ by the resolution of the 14~bit ADC.

Frame-transfer CCDs are suitable for space missions as they eliminate the need for mechanical shutters, which are prone to operational lifetime issues. A major inherent drawback of this architecture is charge smearing \citep{Powell1999, Ruyten1999, Knox2007, Feng2013}. Standard practice to mitigate this artifact involves maintaining an exposure time much longer than the frame transfer time. In the case of PolCam, the impact of charge smearing was not fully anticipated during the design phase. The instrument operates in a regime where the frame transfer time exceeds the exposure time, resulting in severe smear artifacts.

The CCD electronics were designed to read out only six specific lines (one for each channel) from the $1024\times1024$ full-frame array (i.e., six lines acquired per frame). This configuration utilizes the frame-transfer CCD to emulate six line-scan cameras performing push-broom observations. Each line comprises 1024~active pixels, flanked by 16~dark reference pixels on both the left and right ends. During readout, 8~blank elements are appended to each side, as indicated in the bottom row of Figure~\ref{fig:ccd}(c), and 4~dummy pixels are added to each side after readout. The transmitted telemetry data consist of 1080~pixels ($= 1024~\mathrm{active} + 32~\mathrm{dark} + 16~\mathrm{blank} + 8~\mathrm{dummy}$ pixels) per line.

\section{Image Artifact Correction}\label{sec:inscal}

Radiometric calibration generally entails the conversion of raw digital numbers into radiance. For PolCam, this process is modeled by the following equation:
\begin{equation}\label{eq:radio}
L = \frac{DN - Dark - Smear}{  Flat \times t \times T^\mathrm{o} \times T^\mathrm{b} \times T^\mathrm{p} \times R }
\end{equation}
where $L$ is the radiance in units of $\mathrm{W\,m^{-2}\,{\mu}m^{-1}\,sr^{-1}}$, $DN$ is the raw digital number in units of Analog-to-Digital Units (ADU), $t$ is the exposure time, and $T^\mathrm{o}$, $T^\mathrm{b}$, and $T^\mathrm{p}$ denote the transmittances of the optics, bandpass filters, and polarizers, respectively. $R$ is the responsivity of the CCD, relating the $DN$ per exposure time to radiance and incorporating both system gain and quantum efficiency. These parameters depend on wavelength, and their product is expressed as the total optical efficiency ($\eta = T^\mathrm{o} \times T^\mathrm{b} \times T^\mathrm{p} \times R$).

PolCam's primary scientific data product, the degree of linear polarization (DoLP), is derived after applying corrections for dark current, smear artifacts, flat-field non-uniformity, and total optical efficiency. As the absence of an onboard calibrator precludes autonomous efficiency correction, addressing this parameter is a complex task that warrants an independent investigation. Accordingly, we confine the scope of the present study to the first three corrections, deferring a detailed discussion on the efficiency correction to Section~\ref{sec:dis_dolp}. Particular emphasis is placed on mitigating the smear artifacts induced by the frame-transfer CCD, as they constitute a critical source of error in polarimetric measurements.

% T1: dark ---------------------------------------------------
\begin{deluxetable}{cccccccc}
\centering\setlength{\tabcolsep}{4pt}
\tablecaption{Mean and standard deviation of dark current levels measured during the $\sim$2-year nominal mission phase.\label{tab:dark}}
\tablehead{
 & & \colhead{Ch.~1} & \colhead{Ch.~2} & \colhead{Ch.~3} & \colhead{Ch.~4} & \colhead{Ch.~5} & \colhead{Ch.~6}
}
\startdata
\multirow{2}{*}{Left}  & $\mu$    &       837.4 & \phm{0}949.3 & \phm{0}997.0 &       1011.0 &       1022.6 &       1028.2 \\
                       & $\sigma$ & \phm{00}9.1 & \phm{00}11.1 & \phm{00}15.2 & \phm{00}17.0 & \phm{00}16.0 & \phm{00}15.9 \\
\hline
\multirow{2}{*}{Right} & $\mu$    &       968.5 &       1006.2 &       1013.3 &       1015.8 &       1018.3 &       1019.7 \\
                       & $\sigma$ & \phm{00}8.6 & \phm{000}8.8 & \phm{000}9.5 & \phm{00}10.0 & \phm{00}10.5 & \phm{00}10.8 \\
\enddata
\tablecomments{$\mu$ and $\sigma$ are mean and standard deviation in unit of ADU, respectively.}
\vspace*{-6mm}
\end{deluxetable}
% ------------------------------------------------------------

\subsection{Dark Current Removal}\label{sec:dark}

As mentioned in Section~\ref{sec:ccd}, each readout line contains 16~dark reference pixels on both the left and right margins (Figure~\ref{fig:ccd}(c)). The manufacturer recommends using only the middle 8~pixels of the dark reference columns to avoid edge effects. Table~\ref{tab:dark} presents the mean and standard deviation of the dark current levels measured during $\sim$2-year nominal mission phase, separated into left and right columns. Note that the mean value incorporates an electrical charge offset (bias) applied to prevent CCD nonlinearity at low signal levels. While the standard deviation over the two-year period appears relatively high due to CCD temperature variations induced by the Sun's position, the short-term variation ($<$24~min) within a single orbit exhibits a standard deviation of only $\sim$7~ADU. This corresponds to less than 1\% of the typical signal, indicating that the dark level remains stable during an observation sequence.

Several attempts to measure in-flight dark current during the commissioning and early nominal mission phases, via observations of deep space or the lunar nightside, were unsuccessful. At the time, PolCam malfunctioned frequently. Initial operational inexperience across the ground station, spacecraft, and payload teams introduced various procedural and human errors. We focused on observing the lunar dayside under viable observing conditions, and reattempted nightside observations in mid-August 2026. In addition, from the entire dataset collected to date, we identified serendipitous deep-space observations resulting from unplanned attitude maneuvers. These data enabled us to examine how the dark current evolved over a four-year period.

% F2: Dark from in-flight measurements ----------------------
\begin{figure*}[t!]
\centering
\includegraphics[width=\textwidth]{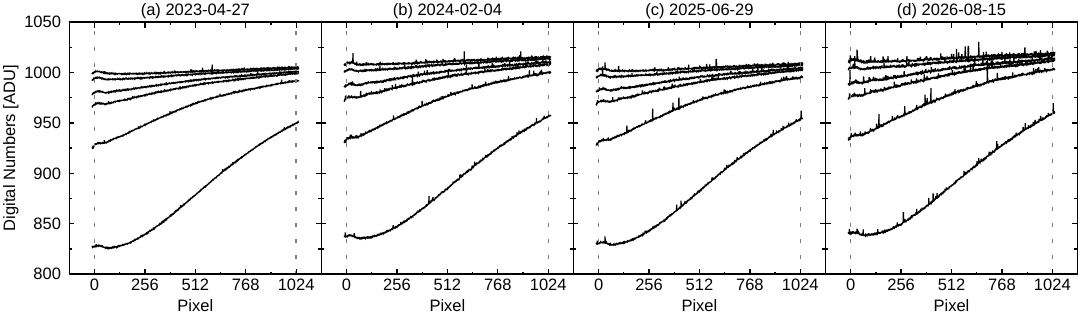}
\caption{Dark current for all six channels (Channels 1--6, from bottom to top), obtained from in-flight measurements. Panels~(a)--(c) show deep-space observations acquired on 2023 April 27, 2024 February 4, and 2025 June 29, respectively, while Panel~(d) shows a nightside observation from 2026 August 15. Pixels with an index below 1 or above 1024 correspond to dark reference pixels.\label{fig:dark}}
\vspace*{5mm}
\end{figure*}
% ------------------------------------------------------------

% F3: Dark reference profile ---------------------------------
\begin{figure}[t!]
\centering
\includegraphics[width=0.47\textwidth]{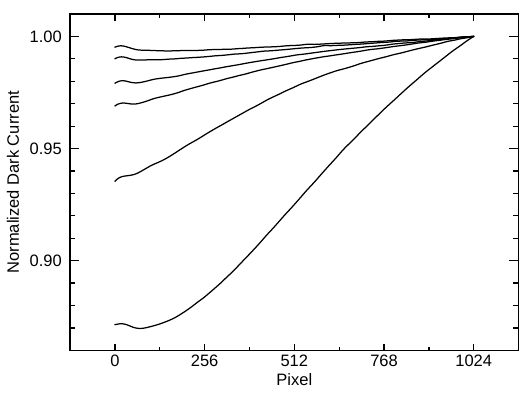}
\caption{Reference profiles of the dark current for all six channels (Channels 1--6, from bottom to top).\label{fig:dark_ref}}
% \vspace{5mm}
\end{figure}
% ------------------------------------------------------------

Figure~\ref{fig:dark} presents the dark current for each channel, derived from observations of deep space (Panels~(a)--(c)) or the lunar nightside (Panel~(d)) and averaged over 21,054 lines per pixel ($\sim$9~min 40~s of observation). The 1D dark current profiles have remained stable since the onset of the nominal mission phase, despite a growing number of warm pixels that exhibit elevated signal levels relative to their surroundings. The profile difference between Figures~\ref{fig:dark}(a) and~\ref{fig:dark}(d) is ($10.0 \pm 1.6$, $9.7 \pm 1.8$, $10.8 \pm 1.5$, $11.6 \pm 1.3$, $12.4 \pm 1.4$, $13.1 \pm 1.3$) ADU for each channel. Standard deviations of all channels are below 2~ADU, driven by warm pixels. Removing these warm pixels reduces the standard deviations to below 1~ADU, confirming that the profile shape has remained constant over the four-year period. The profile shape remains consistent during a single orbit observation, although the overall baseline level increases. For a 10-minute observation, the dark current increased by ($15.3 \pm 0.5$, $13.7 \pm 0.6$, $13.6 \pm 0.9$, $13.7 \pm 0.9$, $14.1 \pm 0.9$, $14.8 \pm 0.7$) ADU across each channel, and for a 20-minute observation, it increased by ($20.6 \pm 0.7$, $18.8 \pm 0.8$, $18.8 \pm 1.1$, $19.1 \pm 1.2$, $19.9 \pm 1.3$, $21.0 \pm 1.2$) ADU.

To estimate the dark level for each pixel within a readout line, we employ the following simple expression:
\begin{equation}
D(x) = D_\mathrm{L} + \frac{f(x) - f(0)}{f(1023) - f(0)} \times (D_\mathrm{R} - D_\mathrm{L})
\end{equation}
where $D_\mathrm{L}$ and $D_\mathrm{R}$ denote the average dark reference values for the left and right margins, respectively. The term $f(x)$ represents the in-flight measured dark current profile smoothed to suppress warm pixels, derived independently for each channel (see Figure~\ref{fig:dark_ref}). $f(0)$ and $f(1023)$ correspond to the leftmost and rightmost boundary values of the profile, respectively. The long-term impacts induced by the suppressed warm pixels are corrected via a flat-fielding process.

% F4: Raw > De-smeared > Flat-fielded (RGS, Crater) ----------
\begin{figure*}[t!]
\centering
\includegraphics[width=\textwidth]{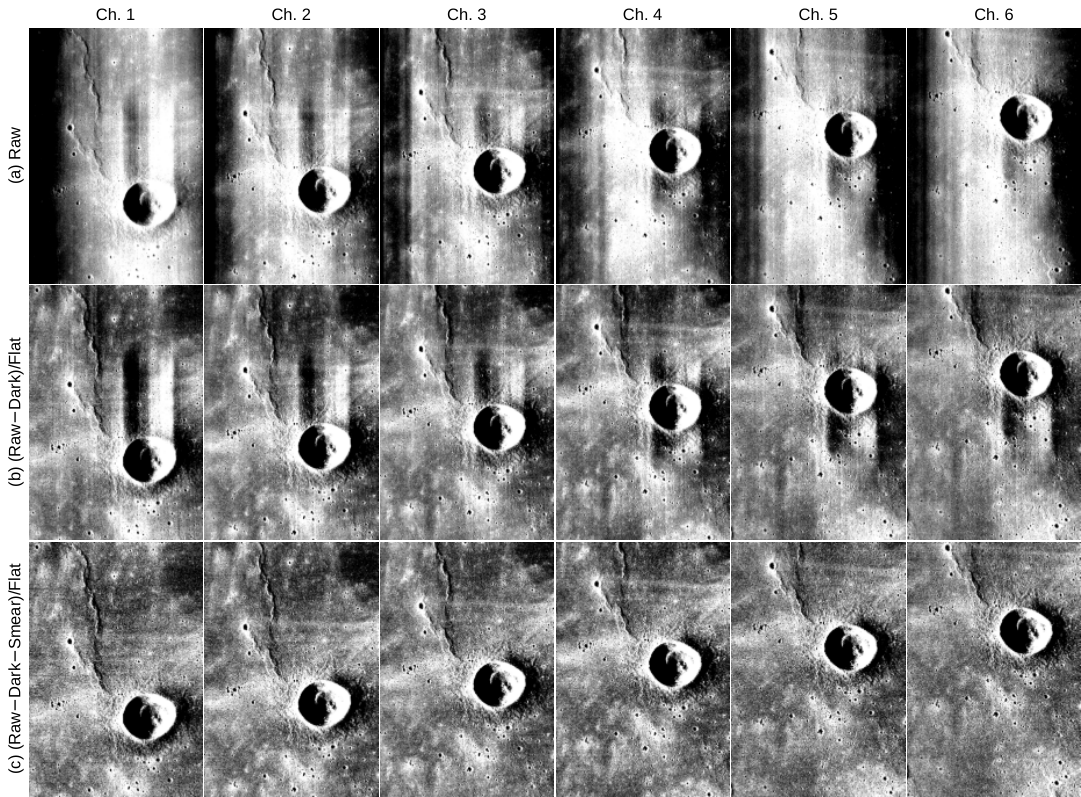}
\caption{Example of severe smear artifacts observed near the Marius~B crater at a phase angle of $\sim$$105^{\circ}$. The rows show: (top) raw; (middle) dark- and flat-corrected; and (bottom) dark-, smear-, and flat-corrected data. Additional examples of smear correction are provided in Appendix~\ref{sec:smear_ex}.\label{fig:smear}}
\vspace*{5mm}
\end{figure*}
% ------------------------------------------------------------

\subsection{Smear Correction}\label{sec:smear}

PolCam raw data exhibit vertical smear artifacts along the orbit direction (see Figure~\ref{fig:smear}(a)). While such smearing is an inherent characteristic of frame-transfer CCDs, an oversight during the detector selection and design phases led to the acquisition of raw data contaminated by these artifacts. As noted in Section~\ref{sec:ccd}, charge smearing arises from the continuous accumulation of charges in the image section during the frame transfer process. To minimize this effect, frame-transfer CCDs are employed in regimes where the exposure time is much longer than the frame transfer time. For PolCam, the frame transfer time ($18.750~\mathrm{ms}$) is more than three times longer than the exposure time ($5.644~\mathrm{ms}$). Accordingly, smear signals are unavoidably prominent.

Although smear signals are distributed across the entire image, they appear in high-contrast regions such as craters observed at high phase angles. Figure~\ref{fig:smear}(a) presents raw images of the Marius~B crater acquired at a phase angle of $\sim$$105^{\circ}$, which reveal smear artifacts extending vertically above and below the crater. These artifacts become even more pronounced in Figure~\ref{fig:smear}(b) after dark and flat-field corrections. Notably, the vertical position of the smear artifacts varies depending on the channel. This shift arises because each channel corresponds to a different physical location (row) on the $1024 \times 1024$ CCD array.

To correct for the smear artifacts generated during readout of PolCam's frame-transfer CCD under highly variable, non-periodic illumination conditions, we adopted the mathematical model proposed by \citetalias{Iglesias2015}. To apply this method, a full-frame image of the entire CCD is required for every frame. Although PolCam records only 6 lines per frame, the full-frame image must be reconstructed for each frame; the detailed procedure is described in Section~\ref{sec:scene}.

The model described in \citetalias{Iglesias2015} defines the smeared signal $\hat{Y}^k_m$ acquired in row $m$ of frame $k$ through the following fundamental equation:
\begin{eqnarray}\label{eq:smear}
\hat{Y}^k_m - \hat{D}_m =\ &r_1& t_t \sum_{j=m+1}^{M-1} g_j S^k_j + g_m t_e S^k_m \\
&+& \frac{1}{2} g_m t_s (S^k_m + S^{k+1}_m) + r_2 t_t \sum_{j=0}^{m-1} g_j S^{k+1}_j ,\nonumber
\end{eqnarray}
where $\hat{D}_m$ is the sum of the dark current and the bias level; $t_t$, $t_e$, and $t_s$ denote the single-pixel charge-transfer time, exposure time, and illumination switching time, respectively; and $r_1$ and $r_2$ are ``ad hoc'' tuning coefficients. The parameters $g$ and $S$ represent the pixel gain and incident photon flux, respectively. Note that, in Equations~(\ref{eq:smear})--(\ref{eq:5}), $m$ is the row index and column index is omitted because these equations hold independently for each column of the image.

Although \citetalias{Iglesias2015} provides a more detailed formulation of $\hat{D}_m$ to account for temporal variations during an observation, this effect is considered negligible for PolCam given its extremely short single-frame integration time ($<$$30~\mathrm{ms}$). Applying a dark correction and simplifying the variables yields:
\begin{align}\begin{split}
\hat{Y}^k_m - \hat{D}_m \rightarrow \hat{Y}^k_m , \quad Y^k_m = g_m t_e S^k_m , \\
\alpha = \frac{t_s}{2t_e}, \quad \delta_1 = \frac{r_1 t_t}{t_e}, \quad \delta_2 = \frac{r_2 t_t}{t_e}.
\end{split}\end{align}
With these substitutions, Equation~(\ref{eq:smear}) can be generalized into a matrix formulation for an entire sensor column:
\begin{equation}
\hat{Y}^k = \mathbf{A} Y^{k} + \mathbf{B} Y^{k+1} ,
\end{equation}
where $\hat{Y}^k$ is the column vector of the measured smeared signals, while $Y^k$ and $Y^{k+1}$ represent the true unsmeared signals of the current and subsequent frames. The transformation matrices $\mathbf{A}$ (upper triangular matrix; diagonal $\alpha$, off-diagonal $\delta_1$) and $\mathbf{B}$ (lower triangular matrix; diagonal $1+\alpha$, off-diagonal $\delta_2$) are derived strictly from the instrument's temporal parameters.

Under non-periodic illumination conditions, recovering the unsmeared images requires a backward recursive inversion process starting from a known or estimated final frame condition, $Y^{K}$. The sequence is restored in the reverse direction using the general solution:
\begin{equation}\label{eq:5}
Y^k = \sum_{j=1}^{K-k} \mathbf{H}^{K-k-j} \mathbf{A}^{-1} \hat{Y}^{K-j} + \mathbf{H}^{K-k} Y^K ,
\end{equation}
where $\mathbf{H}^k = (-\mathbf{A}^{-1}\mathbf{B})^k$. Because the term $\mathbf{H}^k$ acts as a highly damped matrix, this recursive restoration algorithm effectively suppresses error propagation across consecutive frames. It guarantees stable convergence without noise divergence, thereby ensuring the radiometric integrity and inter-channel consistency required for precise polarimetric measurements.

% T2: (r1, r2) -----------------------------------------------
\begin{deluxetable*}{cccccccc}
\setlength{\tabcolsep}{4pt}
\tablecaption{Optimized de-smearing coefficients for each channel.\label{tab:smear}}
\tablehead{
\colhead{Facing} & \colhead{Coeff.} & \colhead{Ch.~1} & \colhead{Ch.~2} & \colhead{Ch.~3} & \colhead{Ch.~4} & \colhead{Ch.~5} & \colhead{Ch.~6}
}
\startdata
\multirow{2}{*}{West} & $r_1$ & $0.33$ & $0.33$ & $0.37$ & $0.35$ & $0.31$ & $0.26$ \\
                      & $r_2$ & $0.36$ & $0.31$ & $0.37$ & $0.32$ & $0.29$ & $0.29$ \\
\hline
\multirow{2}{*}{East} & $r_1$ & $0.40$ & $0.49$ & $0.51$ & $0.55$ & $0.53$ & $0.53$ \\
                      & $r_2$ & $0.34$ & $0.49$ & $0.47$ & $0.52$ & $0.48$ & $0.43$
\enddata
\tablecomments{Facing west: ascending orbits and descending orbits with a yaw-flip. Facing east: descending orbits and ascending orbits with a yaw-flip.}
% \tablecomments{The camera orientation depends on the orbital phase and spacecraft attitude. ``Facing West'' corresponds to ascending orbits in the nominal attitude and descending orbits with a yaw-flip. ``Facing East'' corresponds to descending orbits in the nominal attitude and ascending orbits with a yaw-flip.}
\vspace{-6mm}
\end{deluxetable*}
% ------------------------------------------------------------

As shown in Figure~\ref{fig:smear}(c), the smear artifacts are mitigated to a visually imperceptible level. Additional examples of the smear correction, showing the same Marius~B crater imaged on different dates and at various phase angles, are provided in Appendix~\ref{sec:smear_ex}. Figure~\ref{fig:smear_ex1} presents typical cases in which the smear correction performs effectively, similar to Figures~\ref{fig:smear}(b) and~\ref{fig:smear}(c), whereas Figure~\ref{fig:smear_ex2} shows the poorest-performing cases. During the nominal mission phase, Marius~B crater was observed a total of 10 times: seven were typical cases, and the remaining three were poor cases. All of the poor cases occurred at phase angles exceeding $90^{\circ}$. Severe artifacts in regions other than Marius~B also occur at phase angles greater than $90^{\circ}$. This does not mean all of the smear-corrected data are inaccurate at high phase angles, but it indicates that high-phase-angle data should be used with caution.

This performance relies on the optimization of the two ``ad hoc'' coefficients ($r_1$ and $r_2$) defined in the model of \citetalias{Iglesias2015}. These coefficients compensate for differences in photocharge-generation efficiency between static and transferring states, mitigate synchronization errors during CCD readout, and account for other operational variations. We performed an independent optimization of these coefficients for each channel according to PolCam's facing orientation (west or east), as detailed in Table~\ref{tab:smear}. The separation of east- and west-facing orientations is discussed in Section~\ref{sec:dis_orien}. Channel independence was required because each channel operates with a distinct physical filter and wavelength, resulting in differential response characteristics.

The optimization of these coefficients ($r_1$ and $r_2$) was conducted using a brute-force approach, evaluating the smear correction results for all possible pairs generated from a parameter space ranging from 0.05 to 0.95 at intervals of 0.01. According to \citet{Ruyten1999}, the photocharge-generation efficiency during the frame-transfer process is lower than during the static exposure phase. It follows that obtaining coefficients less than unity is physically well-justified. We selected the optimal values that maximized smear mitigation while minimizing any adverse impact on the degree of polarization.

\subsubsection{CCD Full-Frame Synthesis}\label{sec:scene}

As mentioned in Section~\ref{sec:intro}, the de-smearing model described in \citetalias{Iglesias2015} is theoretically applicable to PolCam. There are, however, two major differences: their study demonstrated the application to periodic illumination within a fixed field of view using full-frame readout, whereas PolCam operates under non-periodic illumination, transmitting only 6 of the 1024~lines. To compensate for the missing signal in the unread 1018~lines, we reconstructed the full-frame data based on the recorded 6~lines. This reconstruction utilized the satellite's orbital speed, pixel pointing vectors, and altitude to map the recorded lines onto the full CCD geometry. We incorporated the ground-measured full-frame efficiency to normalize the relative sensitivity between the recorded and unread lines. For each exposure instance, we synthesized the $1024 \times 1024$ scene projected onto the CCD image section and applied the de-smearing model to this reconstructed full-frame data.

To reconstruct the instantaneous frame that a full \(1024 \times 1024\) CCD would have observed at a given timestamp, the procedure relies on the three-dimensional position vector of the intercept point of each pixel's line of sight with the lunar surface. For a CCD pixel $(x, y)$, the intercept point is expressed in terms of the surface longitude $\lambda^{k}(x,y)$, latitude $\phi^{k}(x,y)$, and radius $r^{k}(x,y)$ at the $k$-th timestamp. The index $k$ represents both the observation frame index and the row index in the image.

We first prepare two position vectors for the $k$ and $k+1$ observation frames, and then compute two types of displacement: frame-to-frame and inter-row. The components of the frame-to-frame displacement vector ($\boldsymbol{d}^\mathrm{f}(x,y) = \left[ d^\mathrm{f}_\mathrm{c}(x,y), d^\mathrm{f}_\mathrm{a}(x,y) \right]$; the superscript $\mathrm{f}$ stands for frame) are expressed:
\begin{eqnarray}
d^\mathrm{f}_\mathrm{c} &=& \left[ \lambda^{k+1}(x,y) - \lambda^{k}(x,y) \right] r^{k}(x,y) \cos\phi^{k}(x,y)\quad\quad \nonumber\\
d^\mathrm{f}_\mathrm{a} &=& \left[ \phi^{k+1}(x,y) - \phi^{k}(x,y) \right] r^{k}(x,y)\,,
\end{eqnarray}
where the subscripts $\mathrm{c}$ and $\mathrm{a}$ denote the direction (cross- and along-track). This frame displacement captures how far the ground footprint of pixel $(x, y)$ moves between the $k$ and $k+1$ timestamps due to spacecraft motion. The components of the inter-row displacement vector ($\boldsymbol{d}^\mathrm{r}(x,y) = \left[ d^\mathrm{r}_\mathrm{c}(x,y), d^\mathrm{r}_\mathrm{a}(x,y) \right]$; the superscript $\mathrm{r}$ stands for row) at the $k$-th frame are expressed:
\begin{eqnarray}
d^\mathrm{r}_\mathrm{c} &=& \left[ \lambda^{k}(x,y) - \lambda^{k}(x,y_\mathrm{r}) \right] r^{k}(x,y_\mathrm{r}) \cos\phi^{k}(x,y_\mathrm{r})\quad\quad \nonumber\\
d^\mathrm{r}_\mathrm{a} &=& \left[ \phi^{k}(x,y) - \phi^{k}(x,y_\mathrm{r}) \right] r^{k}(x,y_\mathrm{r})\,,
\end{eqnarray}
where $y_\mathrm{r}$ is the index of the channel's reference row. It sets the reference row position and the corresponding channel data depend on the range of $y$, as given by the following conditional expression:
\begin{equation}\label{eq:yref}
y_\mathrm{r} = \begin{cases} 
\phm{0}48 & \text{if }\ \phm{00}0 \le y < \phm{0}138 \quad (\text{Ch. 1}) \\ 
180 & \text{if }\ 138 \le y < \phm{0}286 \quad (\text{Ch. 2}) \\ 
368 & \text{if }\ 286 \le y < \phm{0}478 \quad (\text{Ch. 3}) \\ 
564 & \text{if }\ 478 \le y < \phm{0}670 \quad (\text{Ch. 4}) \\ 
724 & \text{if }\ 670 \le y < \phm{0}838 \quad (\text{Ch. 5}) \\ 
884 & \text{if }\ 838 \le y < 1024 \quad (\text{Ch. 6})\ .
\end{cases}
\end{equation}
The row displacement captures the static ground offset between row $y$ and the reference row $y_\mathrm{r}$, i.e. the physical along-track separation between the two rows, expressed in the same tangent-plane frame.

The position of the observational data corresponding to the coordinates viewed by a specific CCD pixel can be calculated from the ratio of the row displacement to the frame displacement of the reference row:
\begin{eqnarray}
i &=& x + \frac{d^\mathrm{r}_\mathrm{c}(x,y)}{d^\mathrm{f}_\mathrm{c}(x,y_\mathrm{r})} = x + \frac{\lambda^{k}(x,y) - \lambda^{k}(x,y_\mathrm{r})}{\lambda^{k+1}(x,y_\mathrm{r}) - \lambda^{k}(x,y_\mathrm{r})} \\
j &=& k + \frac{d^\mathrm{r}_\mathrm{a}(x,y)}{d^\mathrm{f}_\mathrm{a}(x,y_\mathrm{r})} = k + \frac{\phi^{k}(x,y) - \phi^{k}(x,y_\mathrm{r})}{\phi^{k+1}(x,y_\mathrm{r}) - \phi^{k}(x,y_\mathrm{r})}\,.\quad\quad
\end{eqnarray}
These equations compute the projected coordinates $(i, j)$ in the observed image for a specific CCD pixel $(x, y)$. Since calculated actual coordinates are real numbers, the corresponding digital numbers are determined via bilinear interpolation. Applying this procedure to all CCD pixels generates a synthesized full-frame image.

% F5: Scene Synthesis -----------------------------------------
\begin{figure*}[p!]
\centering
\includegraphics[width=\textwidth]{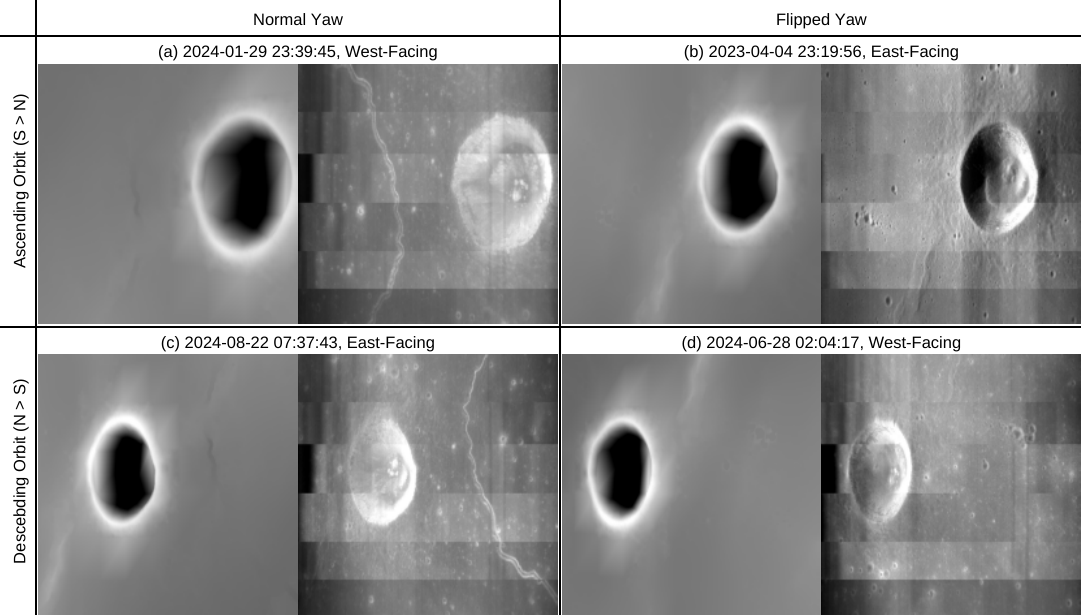}
\caption{Synthesized full-frame images of the same crater observed at different times, generated based on orbit direction (ascending or descending) and yaw-flip status. Each panel shows (left) the expected elevation as projected onto the CCD frame, along with (right) the corresponding synthesized $1024 \times 1024$ CCD image. Labels indicate the observation time and the camera-facing direction (East or West).\label{fig:scene}}
\vspace*{15mm}
% \vspace*{5mm}
% \end{figure*}
% % ------------------------------------------------------------

% % F6: Scene Synthesis -----------------------------------------
% \begin{figure*}[t!]
% \centering
\includegraphics[width=\textwidth]{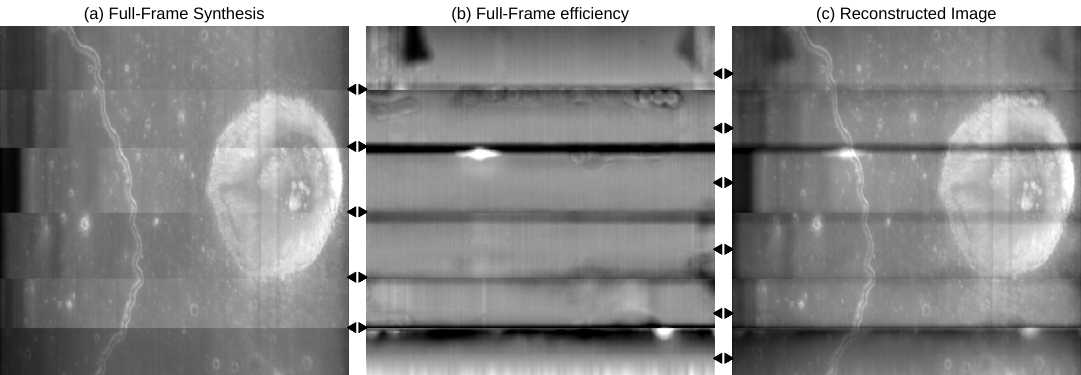}
\caption{Example of a reconstructed image. (a) Synthesized full-frame image, (b) ground-measured full-frame efficiency, (c) reconstructed image, obtained as the product of Panels~(a) and~(b). The triangle symbols between Panels~(a) and~(b) mark the channel boundary locations, and those between Panels~(b) and~(c) indicate the location of each channel. In all Panels, Channels 1--6 are ordered from bottom to top.\label{fig:frame}}
% \vspace*{5mm}
\end{figure*}
% ------------------------------------------------------------

PolCam operates under four possible viewing geometries, depending on the orbit direction and yaw-flip status. The spacecraft travels either south-to-north or north-to-south, and executes a yaw-flip every six months (in late January and early August) to maximize the charging efficiency of its solar panels. Figure~\ref{fig:scene} presents the expected elevation projected onto the CCD frame, alongside the synthesized full-frame CCD images, for all four viewing geometries. The synthesized images align with the expected elevations regardless of viewing geometry. The synthesized full-frame image undergoes a row-by-row relative efficiency correction using ground-measured full-frame efficiency data (see Figure~\ref{fig:frame}). This step completes the reconstructed image, which is utilized for smear correction.

\citetalias{Iglesias2015} used the entire CCD array to observe a fixed field of view, whereas PolCam utilizes only 6 of the 1024 lines to perform push-broom observations that track the orbiter's rapid motion. Smear correction for partial-array readouts, where a subset of rows is recorded, was not considered in \citetalias{Iglesias2015}. This operational constraint is the primary reason that PolCam's smear artifacts cannot be fully corrected, as PolCam cannot measure the actual charge accumulated in the 1018 unread lines during each frame observation. Our smear correction process must rely on values reconstructed from the 6 recorded lines, combined with the ground-measured CCD efficiency. These ground-measured data carry uncertainties, since instrument performance degrades during the launch phase and throughout extended orbital operations. Despite these suboptimal conditions, the proposed algorithm yields smear mitigation sufficient for scientific analysis.

\subsection{Flat-Fielding}\label{sec:flat}

Although channel-specific flat-fields were obtained during pre-launch ground tests \citep{Jeong2023}, they were acquired without accounting for charge smearing effects, rendering them unsuitable for application to on-orbit data. We generated master flats for each channel by integrating columns from the on-orbit data. PolCam's large observation volume, approximately 50,000~lines per orbit, ensures the statistical robustness of the resulting master flats.

During the early phase of operations, we observed multiple flat-field variations that occurred immediately following satellite maneuvers. To track these temporal variations, we monitored daily flat-fields, and constructed distinct master flats for intervals of approximately one month. These intervals were defined by key events such as attitude maneuvers, orbit corrections, yaw-flips, and abrupt flat-field variations.

% F7: Flat variation -----------------------------------------
\begin{figure}[t!]
\centering
\includegraphics[width=0.47\textwidth]{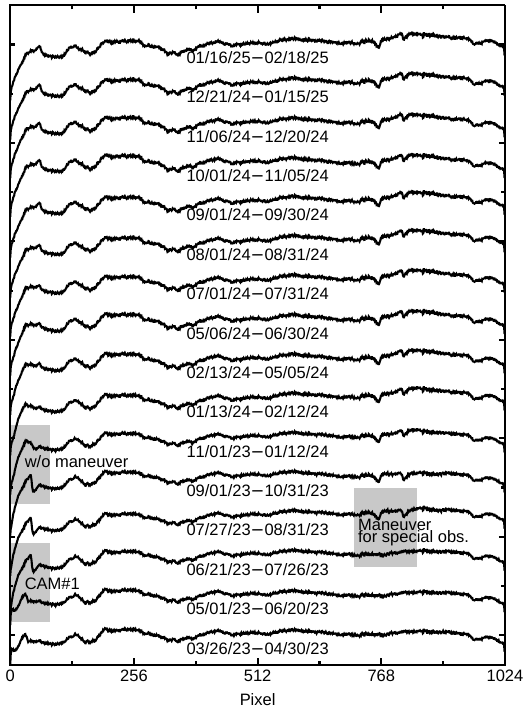}
\caption{Master flats for Ch.~3 during the nominal mission phase. Solid lines represent the master flats generated from on-orbit data acquired during the periods indicated under the each line. Gray shaded boxes highlight flat-field variations that occurred immediately following the collision avoidance maneuver and attitude maneuver. An additional variation was observed without any known maneuver.\label{fig:flat}}
% \vspace*{5mm}
\end{figure}
% ------------------------------------------------------------

Figure~\ref{fig:flat} presents the sequence of master flats for Ch.~3 derived throughout the nominal mission phase. Similar variations are observed across the other channels. The first notable variation occurred immediately following the collision avoidance maneuver (CAM) for Chandrayaan-3 on 2023 June~21, at 01:58:09~UTC \citep{Song2026}, resulting in a localized decrease in incident intensity into the CCD near the 50th pixel. The second variation followed an attitude maneuver executed for the special observation of Humboldt crater on 2023 July~26 at 20:08:47~UTC, introducing a ``eyetooth-shaped'' feature near the 780th pixel. The third variation occurred on 2023 October~31, reappearing near the 50th pixel, although this event was not correlated with any recorded maneuver.

A comparison of the images before and after flat-fielding is presented in the top and middle rows of Figure~\ref{fig:smear}. The raw data (top row) exhibit vertical stripes and intensity fall-off (vignetting) toward the left and right edges. In contrast, the corrected images demonstrate that these artifacts have been effectively mitigated, restoring overall radiometric uniformity.

\section{Degree of Linear Polarization}\label{sec:dolp}

This section presents the degree of linear polarization (DoLP) measurements, constituting the final data product of this study and the primary scientific objective of PolCam. Deriving these measurements requires the image artifact correction described in Section~\ref{sec:inscal} and the precise co-registration of all channels using the 3D coordinates established via our geometric calibration \citepalias{Baek2025}. The mathematical derivation for calculating the DoLP using the three distinct polarizer angles is detailed in Appendix~\ref{sec:calcdop}. The DoLP values were computed assuming uniform efficiency across all channels and an ideal extinction ratio (Equation~(\ref{eq:P430})). Further discussion concerning the implications of channel-specific efficiencies, including non-ideal polarizer transmittances, is deferred to Section~\ref{sec:dis_dolp}.

% F8: DoLP before and after de-smearing -----------------------
\begin{figure}[t!]
\centering
\includegraphics[width=0.47\textwidth]{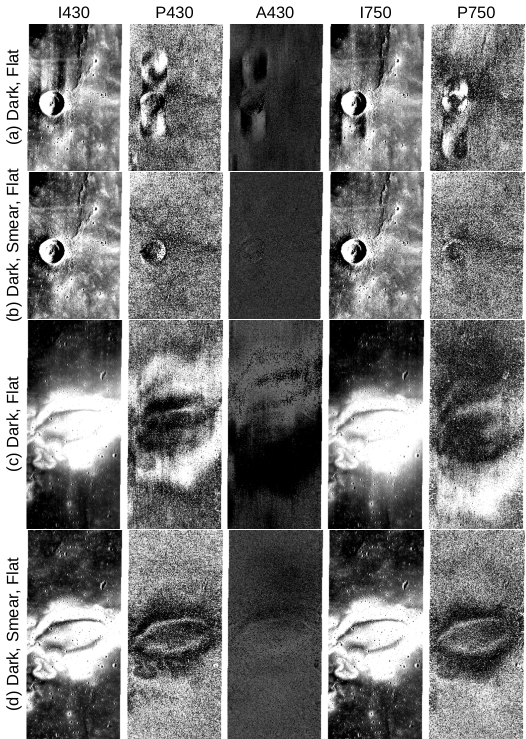}
\caption{Comparison of high-contrast regions before and after smear correction. Panels~(a) and~(b) show the Marius~B crater, while Panels~(c) and~(d) display the eye-shaped feature of the Reiner Gamma swirl. From left to right, the columns correspond to: 430~nm Intensity (I430), 430~nm DoLP (P430), 430~nm AoLP (A430), 750~nm Intensity (I750), and 750~nm DoLP (P750).\label{fig:dop}}
% \vspace*{5mm}
\end{figure}
% ------------------------------------------------------------

Figure~\ref{fig:dop} presents the DoLP images before and after smear correction. Two representative regions were selected where charge smearing effects are pronounced: the partially shadowed Marius~B crater and the Reiner Gamma swirl, which exhibits high albedo contrast against its surroundings. Prior to correction, the polarization signals are severely contaminated by charge smearing, making it difficult to identify surface features (Figures~\ref{fig:dop}(a) and~\ref{fig:dop}(c)). After correction, the polarimetric features are clearly resolved (Figures~\ref{fig:dop}(b) and~\ref{fig:dop}(d)) and exhibit the expected inverse relationship with brightness \citep[Umov's law;][]{Umov1905, Dollfus1971, Wolff1980, Shkuratov1992}.

% F9: DoLP map ------------------------------------------------
\begin{figure*}[t!]
\centering
\includegraphics[width=\textwidth]{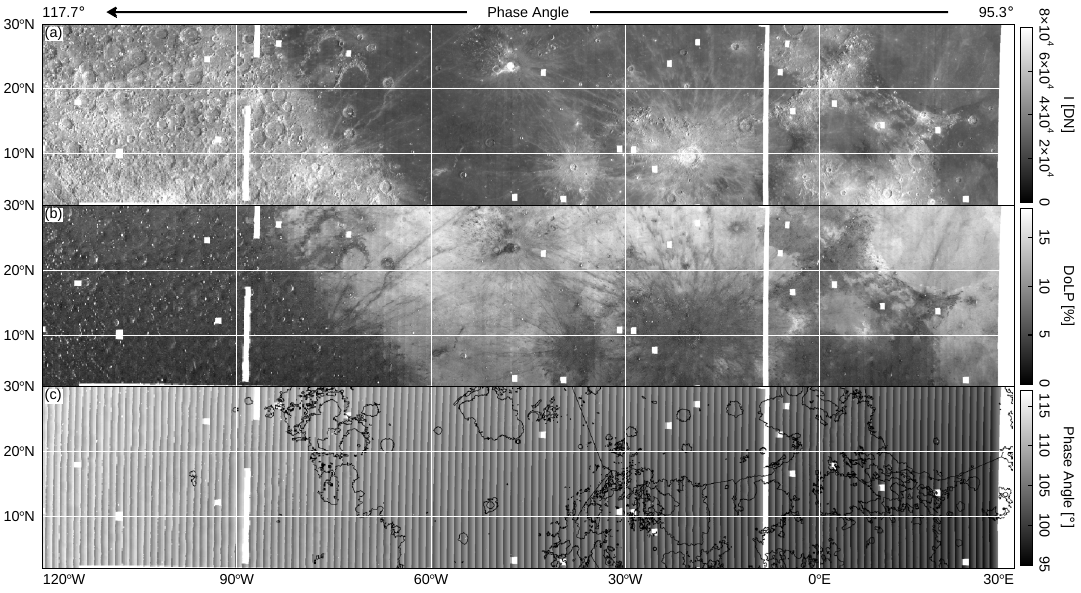}
\caption{Mosaicked 430~nm maps generated from observation data (Channels 2--4) acquired during the early nominal mission phase (2023 March~26 to April~8). The spatial resolution is $128~\mathrm{PPD}$ ($\sim$$240~\mathrm{m/pixel}$). Panels~(a) and~(b) display the Stokes parameter $I$ and DoLP maps, respectively. Panel~(c) shows the corresponding phase angles with the lunar mare boundaries. The average phase angle was $95.^{\!\!\,\circ}3$ during the first orbit (rightmost edge) and progressively increased to $117.^{\!\!\,\circ}7$ throughout the observation period.\label{fig:dopmap}}
\vspace*{3mm}
\end{figure*}
% ------------------------------------------------------------

Figure~\ref{fig:dopmap} presents the mosaicked intensity and DoLP maps (Panels~(a) and~(b), respectively), acquired during the early mission phase from 2023 March~26 to April~8, covering the region from $-120^{\circ}$ to $60^{\circ}$ in longitude and $2^{\circ}$ to $30^{\circ}$ in latitude. Consistent with ground-based observations \citep{Korokhin2005, Shkuratov2007, Jeong2015, Wang2024}, the lunar maria exhibit higher polarization than the highlands. The strong adherence to Umov’s law across a broad area validates that our image artifact correction pipeline enables scientifically meaningful polarimetric measurements.

% F10: DoLP contour ------------------------------------------
\begin{figure}[t!]
\centering\vspace*{10mm}
\includegraphics[width=0.47\textwidth]{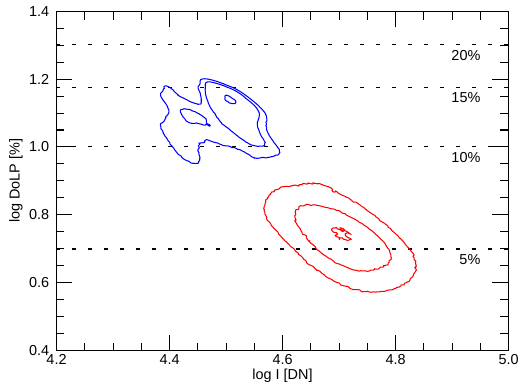}
\caption{Contour plot of the Stokes parameter $I$ versus DoLP at 430~nm on a logarithmic scale, based on the data from Figures~\ref{fig:dopmap}(a) and~\ref{fig:dopmap}(b). The data are distinctly separated into maria (blue contours) and highlands (red contours), with the contours representing the 50th, 75th, and 99th percentiles.\label{fig:contour}}
% \vspace*{5mm}
\end{figure}
% ------------------------------------------------------------

Figure~\ref{fig:contour} displays contour plots of the Stokes parameter $I$ (Figure~\ref{fig:dopmap}(a)) versus DoLP (Figure~\ref{fig:dopmap}(b)). The regions are delineated into maria (blue contours) and highlands (red contours) based on digitized lunar mare boundaries \citep[solid lines in Figure~\ref{fig:dopmap}(c)]{Nelson2014}. These plots illustrate an inverse relationship between intensity and DoLP, consistent with Umov's law. Since the derived DoLP values are lower than those of ground-based observations, the current analysis is limited to a qualitative assessment. This discrepancy is attributed to unresolved channel-specific efficiencies; a detailed discussion on the absolute calibration of DoLP is provided in Section~\ref{sec:dis_dolp}.

\section{Summary and Discussion}\label{sec:condis}

The image artifact correction of PolCam required for accurate polarimetric measurements was completed by implementing dark current removal, smear correction, and flat-fielding (see Section~\ref{sec:inscal}). Integrating these results with our geometric calibration \citepalias{Baek2025}, we derived the polarimetric measurements (see Section~\ref{sec:dolp}). PolCam is ready for the scientific `qualitative' investigations of polarimetric properties at the lunar surface. Recent publications \citep{Shkuratov2026a, Shkuratov2026b, Shkuratov2026c} on irregular mare patches confirm that PolCam data provide complementary insights that support the spectral characteristics observed in existing multi-band datasets.

A primary contribution of this study is the successful mitigation of smear artifacts inherent to the frame-transfer CCD architecture (see Figure~\ref{fig:dop}). Despite the absence of pre-launch ground characterization for charge smearing, we achieved substantial correction using solely on-orbit data. This correction restored scientific data that otherwise might have been unusable for reliable polarization retrieval. Nevertheless, we must remain cautious regarding the potential impacts of any residual smear (see Section~\ref{sec:dis_smear}).

Developed under tight mass, volume, and schedule constraints, PolCam could neither undergo pre-launch testing nor carry an onboard calibration target, leaving on-orbit calibration as the sole viable option. Given the infeasibility of conducting in-flight calibration of the optical system, refining absolute DoLP measurements using exclusively PolCam's on-orbit data is unachievable. The completion of both absolute radiance and photometric calibrations remains a critical prerequisite for enabling quantitative analysis, which will be the focus of the third paper in this series. These remaining challenges and future directions are discussed in Section~\ref{sec:dis_dolp}.

\subsection{Facing-Orientation Dependence of Smear}\label{sec:dis_orien}

During the optimization of smear correction, an unexpected phenomenon was identified: the magnitude of the smear artifacts depends on PolCam's facing orientation. An exhaustive analysis of potential driving variables---including spacecraft attitude, altitude, orbital direction (ascending/descending), orbital speed, and yaw-flip maneuvers---was conducted, but failed to isolate a definitive physical cause. Facing orientation was identified as the primary empirical dependency governing the smear variation, a dependence that is puzzling given the symmetry expected from the observation geometry.

An Investigation was conducted to determine whether this asymmetry could reflect a genuine brightness difference between east- and west-facing observations. Because PolCam's survey spans more than two years, observations at a given phase angle within the range $45^{\circ}$--$135^{\circ}$ are evenly distributed across orbit directions and facing orientations; the effect is not an artifact of an unbalanced local solar time sampling. East- and west-facing observations of the same terrain at a similar phase angle correspond to a mirror-symmetric illumination geometry, under which classical scattering theory predicts both the reflectance and polarization of an azimuthally isotropic surface to be invariant \citep{Hapke2012}. A scene-brightness origin for the effect is unexpected, though the exact azimuthal symmetry upon which this assumption relies has itself been questioned \citep{Shkuratov2012}. Accordingly, a dedicated analysis of PolCam's own reflectance and angle-of-linear-polarization data as a function of scattering-plane orientation will be required to confirm this expectation, rather than by appeal to theory.

The asymmetry is present in the raw images themselves, not only after smear correction. The top rows of Figures~\ref{fig:smear_ex2}(a) and~\ref{fig:smear_ex2}(b) show raw images of the same crater acquired while facing west and east, respectively, under comparable altitudes and orbital speeds. The west-facing data display more elongated and pronounced smear streak than the east-facing data of the same feature. At equivalent orbital speeds, smear length should depend only on the readout row's position, which is fixed for each channel; the two cases should match, revealing an observed discrepancy. This dependence must originate either upstream of image formation, in the illumination or scattering geometry as discussed above, or within the detector and readout electronics themselves. Another potential factor is the fixed orientation of the polarizer axes relative to the scattering plane. We were unable to conclusively isolate the exact mechanism in this study.

\subsection{Residual Smear}\label{sec:dis_smear}

Although we mitigated smear artifacts to a remarkable degree using solely on-orbit data, their complete removal remains unattainable due to a structural limitation unique to PolCam: the system records only 6 of the 1024 CCD lines. The charge accumulated in the unread 1018~lines (the source of the smearing) cannot be directly quantified and must instead be reconstructed. Based on the temporal variations in the flat-field responses (Figure~\ref{fig:flat}), we deduce that PolCam's optical throughput degraded over the mission lifetime. This degradation alters the amount of light incident on each pixel non-uniformly, causing the resulting smear to vary on pixel by pixel. The reconstructed rows cannot account for this pixel-specific optical degradation present in the unread regions, rendering flawless smear removal impossible. Caution is warranted regarding potential residual smear when utilizing data derived from inter-channel differences, including the Stokes parameters ($Q, U$) and DoLP.

Our on-orbit data analysis revealed that the smearing trends differ depending on whether the camera is facing east or west. The correction coefficients ($r_1$ and $r_2$) were independently optimized for these two viewing geometries (Table~\ref{tab:smear}). While this approach yields robust corrections for the majority of the nominal phase data, it cannot be guaranteed that these static values are applicable across all observation dates. Theoretically, the coefficients $r_1$ and $r_2$ represent intrinsic CCD characteristics \citepalias[e.g., photocharge-generation efficiency, synchronization errors, and other operational differences;][]{Iglesias2015}; for PolCam, these coefficients are forced to simultaneously compensate for the unknown optical degradation within the unread 1018~lines. Akin to the temporal variations observed in the flat-field, these coefficients likely fluctuate over time for each channel.

Future research must focus on quantifying the residual smear and tracking its temporal variations to further characterize $r_1$ and $r_2$. The brightness of the lunar surface varies substantially with topography, incidence angle, and emission angle, which makes it difficult to define the magnitude of residual smear and distinguish it from the surrounding signal. We expect that quantification will become feasible once photometric calibration allows normalization for incidence and emission angle. This refinement will be critical for ensuring the accuracy of the absolute DoLP calibration and the subsequent derivation of maximum polarization via phase curve fitting.

\subsection{Degree of Linear Polarization}\label{sec:dis_dolp}

As shown in Figure~\ref{fig:contour}, the 430~nm observations yield DoLP values of $10$--$15\%$ for the maria and $4$--$7\%$ for the highlands. These values are $\sim$$5\%$ lower than the maximum DoLP measurements (maria: $15.8$--$22.4\%$, highlands: $7.1$--$10.0\%$) reported by \citet{Jeong2015} from ground-based observations in the comparable B-band ($\lambda_\mathrm{eff} = 443.5~\mathrm{nm}$). Even when considering that Figure~\ref{fig:contour} presents data across a phase angle range ($95^{\circ}$--$115^{\circ}$) that encompasses the maximum polarization, rather than isolating the exact maximum peaks, our derived DoLP remains noticeably lower than the ground-based results.

Validating absolute DoLP values requires correcting for the total optical efficiency of each channel. Without this correction, only a qualitative analysis of the DoLP is feasible (see Section~\ref{sec:dolp}). Given the absence of an onboard calibration target for PolCam, estimating these efficiencies for each channel relies on cross-calibration with other orbiters or ground-based observations. This process, in turn, requires a comprehensive photometric calibration.

We aim to verify whether uncalibrated differences in total optical efficiency among the channels are responsible for the lower-than-expected DoLP. To investigate this, we utilized observational data from Channels~2--4 (430~nm). While calculating the absolute Stokes parameters ($I$, $Q$, and $U$; Equations~(\ref{eq:I430_2})--(\ref{eq:A430_2})) requires estimating the total optical efficiencies for all three channels, computing their ratios (e.g., $Q/I$, $U/I$, and $U/Q$) requires their relative efficiencies, reducing the number of unknown variables to two. The theoretical angle of linear polarization (AoLP), calculated from the observation geometry, was established as a reference. By minimizing the residual between the theoretical and the observed AoLP (Equation~(\ref{eq:A430_2})) via a least-squares approach, we estimated the relative efficiencies among the three 430~nm channels.

% F11: DoLP vs DoLP contour ----------------------------------
\begin{figure}[t!]
\centering
\includegraphics[width=0.47\textwidth]{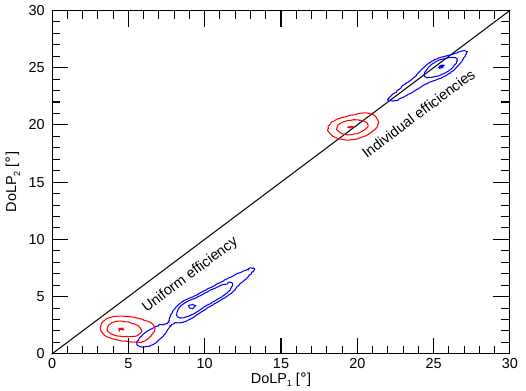}
\caption{Contour plot comparing the DoLP acquired during two different observation periods at similar phase angles. The data are distinctly separated into maria (blue contours) and highlands (red contours), with contours drawn at the 50th, 75th, and 99th percentiles. The contours in the lower-left corner represent values derived under the assumption of uniform efficiency across the CCD, whereas the upper-right contours incorporate the relative efficiency differences between the channels.\label{fig:PvsP}}
% \vspace*{5mm}
\end{figure}
% ------------------------------------------------------------

Figure~\ref{fig:PvsP} presents a contour plot comparing the DoLP acquired during two distinct observation periods at similar phase angles ($95^{\circ}$--$115^{\circ}$). The data for Period~1 ($x$-axis), which corresponds to Figure~\ref{fig:dopmap}(b), were obtained with an east-facing orientation. The Period~2 data ($y$-axis), collected between 2024 May~28 and June~6, cover the overlapping area ($-120^{\circ}$--$0^{\circ}$ in longitude and $2^{\circ}$--$3^{\circ}$ in latitude) but were acquired with a west-facing orientation. The contours in the lower-left corner of Figure~\ref{fig:PvsP} represent the DoLP calculated under the assumption of an ideal optical system (i.e., uniform total optical efficiency across all channels; Equation~(\ref{eq:P430}) with an ideal extinction ratio, $T^{\perp} \sim 0$). Under this idealized assumption, despite observing the same region at similar phase angles, the measured DoLP distributions differ noticeably between the two periods and remain anomalously low (below $15\%$ for Period~1 and below $10\%$ for Period~2).

To rectify this discrepancy, the relative efficiencies of the channels were estimated by assuming a baseline efficiency of unity for Channel~4 (the $0^{\circ}$ polarizer angle), closest to the CCD center. Through this approach across the three polarizer orientation angles ($120^{\circ}$, $60^{\circ}$, and $0^{\circ}$), the relative efficiencies $[\eta_{120}, \eta_{60}, \eta_{0}]$ were determined as $[0.724, 0.757, 1.000]$ for Period~1 and $[0.700, 0.713, 1.000]$ for Period~2.

The upper-right contours in Figure~\ref{fig:PvsP} display the DoLP distributions after applying these relative efficiencies. Both the maria (blue contours) and the highlands (red contours) demonstrate a strong linear correlation along the 1:1 line, with their overall DoLP increasing to levels consistent with ground-based observations \citep{Korokhin2005, Shkuratov2007, Jeong2015, Wang2024}. This alignment is logically sound for repeat observations of the same region at identical phase angles, confirming that discrepancies in total optical efficiency among channels are a pivotal factor in absolute DoLP measurements.

Estimating relative efficiencies via the AoLP is advantageous, as it relies solely on PolCam's on-orbit data; however, this method has clear limitations. The estimated efficiencies can fluctuate by up to $\sim$10\% depending on data quality and the selected region. This approach does not resolve the absolute efficiencies of individual channels, nor does it apply to the 320~nm (Ch.~1) and 750~nm (Ch.~5 and~6) bands, where deriving the observed AoLP is unfeasible. Precise cross-calibration with other orbiter missions or ground-based measurements remains indispensable. Reliable comparative analysis and normalization against established references will require a clear understanding of the photometric properties of the PolCam data for comparing PolCam's off-nadir ($\sim$$45^{\circ}$-tilted) observations with legacy datasets from past orbital missions, which primarily acquired nadir observations. The absolute calibration of both the Stokes parameters and the DoLP will need to be established through cross-calibration with previously verified datasets.

% Please use the acknowledgment and contribution environments. This will 
% be anonomyized when the "anonymous" style option is used. 
\begin{acknowledgments}
% The work by SSK was supported by the Korea Astronomy and Space Science Institute under the R\&D program (2023-1-850-09) supervised by the Ministry of Science and ICT.
The constructive comments and thorough suggestions provided by the anonymous referee are gratefully acknowledged.
The work by SSK was supported by the Korea Astronomy and Space Science Institute under the R\&D Program (2026-185100) supervised by the Ministry of Science and ICT.
\end{acknowledgments}

\vspace{2mm}
\begin{contribution}
% Option 1: Standard Narrative (Most Common)
% Option 2: CRediT Format (Structured)
\textbf{Kilho Baek}: Data curation, Formal analysis, Methodology, Investigation, Software, Validation, Visualization, and Writing --- original draft.
\textbf{Sungsoo S. Kim}: Conceptualization, Project administration, Funding acquisition, Resources, Supervision, and Writing --- review \& editing.
\textbf{Minsup Jeong}: Formal analysis, Project administration, and Writing --- review \& editing.
\textbf{Young-Jun Choi}: Project administration.
\end{contribution}


\clearpage
\appendix
\restartappendixnumbering

\section{Polarization Formula with Three Polarizers}\label{sec:calcdop}

The Stokes vector quantitatively describes the polarization state of light, while the Mueller matrix characterizes its transformation through an optical element. The resulting polarization state of the outgoing light is expressed by the linear transformation:
\begin{eqnarray}\label{eq:stokes}
S\phm{'} &=& \left[ I\phm{'} Q\phm{'} U\phm{'} V\phm{'} \right]^T \nonumber\\
S' &=& \left[ I' Q' U' V' \right]^T \\
S' &=& M S \nonumber
\end{eqnarray}
where $S$ and $S'$ denote the Stokes vectors of incoming and outgoing light, respectively, and $M$ is the $4 \times 4$ Mueller matrix of the optical system. For a linear polarizer, the generalized Mueller matrix is defined as:
{\small\begin{equation}\label{eq:mueller}
M =
\frac{1}{2}
\left[\begin{array}{cccc}
T^{\parallel} + T^{\perp} & (T^{\parallel} - T^{\perp})\cos{2\Psi} & (T^{\parallel} - T^{\perp})\sin{2\Psi} & 0 \\
(T^{\parallel} - T^{\perp})\cos{2\Psi} & (T^{\parallel} + T^{\perp})\cos^{2}{2\Psi} + 2\sqrt{T^{\parallel} T^{\perp}}\sin^{2}{2\Psi} & (T^{\parallel} + T^{\perp} - 2\sqrt{T^{\parallel} T^{\perp}})\cos{2\Psi}\sin{2\Psi} & 0 \\
(T^{\parallel} - T^{\perp})\sin{2\Psi} & (T^{\parallel} + T^{\perp} - 2\sqrt{T^{\parallel} T^{\perp}})\cos{2\Psi}\sin{2\Psi} & (T^{\parallel} + T^{\perp})\sin^{2}{2\Psi} + 2\sqrt{T^{\parallel} T^{\perp}}\cos^{2}{2\Psi} & 0 \\
0 & 0 & 0 & 2\sqrt{T^{\parallel} T^{\perp}} 
\end{array}\right],
\end{equation}}
where $T^{\parallel}$ and $T^{\perp}$ are the transmittances along the preferred and orthogonal axes, respectively, and $\Psi$ is the orientation angle. For an ideal polarizer, $T^{\parallel} = 1$ and $T^{\perp} = 0$.

Because detectors measure only intensity, extracting the incoming Stokes parameters ($I$, $Q$, and $U$) requires only the first row of Equation~(\ref{eq:mueller}). For observations acquired at three polarizer orientation angles ($\Psi_{1}, \Psi_{2}, \Psi_{3}$), the measured intensities relate to the incoming Stokes parameters:
\begin{equation}\label{eq:mueller_obs}
\left[\begin{array}{c} I'_{1} \\ I'_{2} \\ I'_{3} \end{array}\right] = 
\left[\begin{array}{c}
I^\mathrm{c}_{1}/(T^\mathrm{o}_{1} T^\mathrm{b}_{1} R_{1}) \\
I^\mathrm{c}_{2}/(T^\mathrm{o}_{2} T^\mathrm{b}_{2} R_{2}) \\
I^\mathrm{c}_{3}/(T^\mathrm{o}_{3} T^\mathrm{b}_{3} R_{3})
\end{array}\right] = 
\frac{1}{2}
\left[\begin{array}{ccc}
T^{\parallel}_{1} + T^{\perp}_{1} & (T^{\parallel}_{1} - T^{\perp}_{1})\cos{2\Psi_{1}} & (T^{\parallel}_{1} - T^{\perp}_{1})\sin{2\Psi_{1}} \\
T^{\parallel}_{2} + T^{\perp}_{2} & (T^{\parallel}_{2} - T^{\perp}_{2})\cos{2\Psi_{2}} & (T^{\parallel}_{2} - T^{\perp}_{2})\sin{2\Psi_{2}} \\
T^{\parallel}_{3} + T^{\perp}_{3} & (T^{\parallel}_{3} - T^{\perp}_{3})\cos{2\Psi_{3}} & (T^{\parallel}_{3} - T^{\perp}_{3})\sin{2\Psi_{3}} 
\end{array}\right]
\left[\begin{array}{c} I \\ Q \\ U \\ \end{array}\right],
\end{equation}
where $I'_{n} = I^\mathrm{c}_{n}/(T^\mathrm{o}_{n} T^\mathrm{b}_{n} R_{n})$ is the fully calibrated intensity, and $I^\mathrm{c}_{n}$ is the dark-, flat-, and smear-corrected intensity. The terms $T^\mathrm{o}_{n}$, $T^\mathrm{b}_{n}$, and $R_{n}$ denote the transmittances of the optics, bandpass filter, and CCD responsivity, respectively. The circular polarization component ($V$) is negligible for the lunar surface \citep{Lipskii1967, Shkuratov2011} and is omitted.

\subsection{Shared Filter System}

Conventional ground-based polarimetry typically employs a single set of optics, bandpass filter, and polarizer. Because the orientation angle is changed by rotating the same polarizer, the system parameters ($T^\mathrm{o}$, $T^\mathrm{b}$, $R$, $T^{\parallel}$, $T^{\perp}$) are identical across all measurements, simplifying Equation~(\ref{eq:mueller_obs}) to:
\begin{equation}\label{eq:mueller_obs_shared}
\left[\begin{array}{c}
I^\mathrm{c}_{1} \\
I^\mathrm{c}_{2} \\
I^\mathrm{c}_{3}
\end{array}\right] = 
\frac{T^\mathrm{o} T^\mathrm{b} R}{2}
\left[\begin{array}{ccc}
T^{\parallel} + T^{\perp} & (T^{\parallel} - T^{\perp})\cos{2\Psi} & (T^{\parallel} - T^{\perp})\sin{2\Psi_{1}} \\
T^{\parallel} + T^{\perp} & (T^{\parallel} - T^{\perp})\cos{2\Psi} & (T^{\parallel} - T^{\perp})\sin{2\Psi_{2}} \\
T^{\parallel} + T^{\perp} & (T^{\parallel} - T^{\perp})\cos{2\Psi} & (T^{\parallel} - T^{\perp})\sin{2\Psi_{3}} 
\end{array}\right]
\left[\begin{array}{c} I \\ Q \\ U \\ \end{array}\right].
\end{equation}
A primary advantage of this shared configuration is that when calculating the normalized Stokes parameters ($q = Q/I$ and $u = U/I$) to derive the Degree of Linear Polarization (DoLP), the common radiometric efficiency term ($T^\mathrm{o} T^\mathrm{b} R$) cancels out entirely. The incident polarization state $[I, Q, U]^T$ can be determined by simply inverting the corresponding $3 \times 3$ matrix:
\begin{equation}\label{eq:IQU}
\left[\begin{array}{c} I \\ Q \\ U \\ \end{array}\right] =
\frac{1}{T^\mathrm{o} T^\mathrm{b} R}
\frac{2}{\mathrm{det}\,M}
\left[\begin{array}{c}
(T^{\parallel} - T^{\perp})^{2}\\
{T^{\parallel}}^2 - {T^{\perp}}^2\\
{T^{\parallel}}^2 - {T^{\perp}}^2\\
\end{array}\right]
\left[\begin{array}{ccc}
\sin(2\Psi_{3}-2\Psi_{2}) & \sin(2\Psi_{1}-2\Psi_{3}) & \sin(2\Psi_{2}-2\Psi_{1}) \\
\sin 2\Psi_{2} - \sin 2\Psi_{3} & \sin 2\Psi_{3} - \sin2\Psi_{1} & \sin2\Psi_{1}-\sin 2\Psi_{2} \\
\cos 2\Psi_{3} - \cos 2\Psi_{2} & \cos 2\Psi_{1} - \cos 2\Psi_{3} & \cos 2\Psi_{2} - \cos 2\Psi_{1}
\end{array}\right]
\left[\begin{array}{c} I^\mathrm{c}_{1} \\ I^\mathrm{c}_{2} \\ I^\mathrm{c}_{3} \end{array}\right],
\end{equation}
where $\mathrm{det}\,M = 4 (T^{\parallel} + T^{\perp}) (T^{\parallel} - T^{\perp})^{2} \sin(\Psi_{1}-\Psi_{2}) \sin(\Psi_{2}-\Psi_{3}) \sin(\Psi_{3}-\Psi_{1})$.

Employing the standard orientation angles $[\Psi_{1}, \Psi_{2}, \Psi_{3}] = [120^{\circ}, 60^{\circ}, 0^{\circ}]$ further simplifies the derivation through trigonometric symmetry. Substituting these angles and their corresponding intensities $[I^\mathrm{c}_{120}, I^\mathrm{c}_{60}, I^\mathrm{c}_{0}]$ yields:
\begin{eqnarray}
I &=& \frac{1}{T^\mathrm{o} T^\mathrm{b} R} \frac{1}{T^{\parallel} + T^{\perp}} \frac{2}{3} \left( \phm{2} I^\mathrm{c}_{0} + I^\mathrm{c}_{60} + I^\mathrm{c}_{120} \right), \label{eq:I430}\\
Q &=& \frac{1}{T^\mathrm{o} T^\mathrm{b} R} \frac{1}{T^{\parallel} - T^{\perp}} \frac{2}{3} \left( 2 I^\mathrm{c}_{0} - I^\mathrm{c}_{60} - I^\mathrm{c}_{120} \right), \label{eq:Q430}\\
U &=& \frac{1}{T^\mathrm{o} T^\mathrm{b} R} \frac{1}{T^{\parallel} - T^{\perp}} \frac{2}{\sqrt{3}} \left( I^\mathrm{c}_{60} - I^\mathrm{c}_{120} \right), \label{eq:U430}\\\nonumber\\
% P &=& \sqrt{Q^{2} + U^{2}}\Big/I \nonumber\\
P &=& \frac{T^{\parallel} + T^{\perp}}{T^{\parallel} - T^{\perp}} \frac{2\sqrt{I^\mathrm{c}_{0} ( I^\mathrm{c}_{0} - I^\mathrm{c}_{60} ) + I^\mathrm{c}_{60} ( I^\mathrm{c}_{60} - I^\mathrm{c}_{120} ) + I^\mathrm{c}_{120} ( I^\mathrm{c}_{120} - I^\mathrm{c}_{0} )}}{I^\mathrm{c}_{0} + I^\mathrm{c}_{60} + I^\mathrm{c}_{120}}.\label{eq:P430}
\end{eqnarray}
Equation~(\ref{eq:P430}) is consistent with Equation~(7.10) of \citet{Shkuratov2025} under the assumption of an ideal extinction ratio ($T^{\perp} \sim 0$).

The angle of linear polarization (AoLP; $\Psi_P$) can be directly calculated from $Q$ and $U$. If the AoLP is known a priori, the DoLP can be determined using measurements from only two orientation angles:
\begin{eqnarray}
\Psi_{P} &=& \frac{1}{2} \arctan \frac{U}{Q} = \frac{1}{2} \arctan \frac{\sqrt{3} ( I^\mathrm{c}_{60} - I^\mathrm{c}_{120} )}{2 I^\mathrm{c}_{0} - I^\mathrm{c}_{60} - I^\mathrm{c}_{120}} \label{eq:A430}\\
P &=& \frac{I^\mathrm{c}_{0} - I^\mathrm{c}_{90}}{I^\mathrm{c}_{0} + I^\mathrm{c}_{90}} \frac{1}{\cos(2\Psi_{P})} \label{eq:P750}
\end{eqnarray}

\subsection{Individual Filter System}

In contrast, because PolCam utilizes physically distinct bandpass filters and polarizers for each channel, each row in Equation~(\ref{eq:mueller_obs}) possesses unique $T^{\parallel}$ and $T^{\perp}$ values. Defining the total optical efficiency of each channel as $\eta_{n} = T^\mathrm{o}_{n} T^\mathrm{b}_{n} T^\mathrm{p}_{n} R_{n}$ (Equation~(\ref{eq:radio})), Equation~(\ref{eq:mueller}) can be reformulated for actual data processing as:
\begin{equation}\label{eq:mueller_obs_ind}
\left[\begin{array}{c}
I^\mathrm{c}_{1}/\eta_{1} \\
I^\mathrm{c}_{2}/\eta_{2} \\
I^\mathrm{c}_{3}/\eta_{3} 
\end{array}\right] =
\frac{1}{2}
\left[\begin{array}{ccc}
1 & (T^{\parallel}_{1} - T^{\perp}_{1})/(T^{\parallel}_{1} + T^{\perp}_{1})\cos{2\Psi_{1}} & (T^{\parallel}_{1} - T^{\perp}_{1})/(T^{\parallel}_{1} + T^{\perp}_{1})\sin{2\Psi_{1}} \\
1 & (T^{\parallel}_{2} - T^{\perp}_{2})/(T^{\parallel}_{2} + T^{\perp}_{2})\cos{2\Psi_{2}} & (T^{\parallel}_{2} - T^{\perp}_{2})/(T^{\parallel}_{2} + T^{\perp}_{2})\sin{2\Psi_{2}} \\
1 & (T^{\parallel}_{3} - T^{\perp}_{3})/(T^{\parallel}_{3} + T^{\perp}_{3})\cos{2\Psi_{3}} & (T^{\parallel}_{3} - T^{\perp}_{3})/(T^{\parallel}_{3} + T^{\perp}_{3})\sin{2\Psi_{3}} 
\end{array}\right]
\left[\begin{array}{c} I \\ Q \\ U \\ \end{array}\right].
\end{equation}
As discussed in Section~\ref{sec:dis_dolp}, constraining $\eta_{n}$, $T^{\parallel}_{n}$, and $T^{\perp}_{n}$ across all channels using exclusively on-orbit data is practically impossible. Assuming the use of ideal extinction ratio ($T^{\perp}_{n} \sim 0$) provides the only viable simplification for Equation~(\ref{eq:mueller_obs_ind}):
\begin{equation}\label{eq:mueller_obs2}
\left[\begin{array}{c}
I^\mathrm{c}(\Psi_{1})/\eta_{1} \\
I^\mathrm{c}(\Psi_{2})/\eta_{2} \\
I^\mathrm{c}(\Psi_{3})/\eta_{3} 
\end{array}\right] =
\frac{1}{2}
\left[\begin{array}{ccc}
1 & \cos{2\Psi_{1}} & \sin{2\Psi_{1}} \\
1 & \cos{2\Psi_{2}} & \sin{2\Psi_{2}} \\
1 & \cos{2\Psi_{3}} & \sin{2\Psi_{3}} 
\end{array}\right]
\left[\begin{array}{c} I \\ Q\ \\ U\ \\ \end{array}\right].
\end{equation}
Under these conditions, substituting the specific orientation angles ($0^\circ$, $60^\circ$, and $120^\circ$) provides the expressions for the Stokes parameters ($I, Q, U$), DoLP ($P$), and AoLP ($\Psi_{P}$):
\begin{eqnarray}
I &=& \frac{2}{3} \left( \frac{I^\mathrm{c}_{0}}{\eta_{0}} + \frac{I^\mathrm{c}_{60}}{\eta_{60}} + \frac{I^\mathrm{c}_{120}}{\eta_{120}} \right) \label{eq:I430_2}\\
Q &=& \frac{2}{3} \left( \frac{2 I^\mathrm{c}_{0}}{\eta_{0}} - \frac{I^\mathrm{c}_{60}}{\eta_{60}} - \frac{I^\mathrm{c}_{120}}{\eta_{120}} \right) \label{eq:Q430_2}\\
U &=& \frac{2}{\sqrt{3}} \left( \frac{I^\mathrm{c}_{60}}{\eta_{60}} - \frac{I^\mathrm{c}_{120}}{\eta_{120}} \right) \label{eq:U430_2}\\\nonumber\\
% P &=& \frac{2\sqrt{\frac{I^\mathrm{c}_{0}}{\eta_{0}} ( \frac{I^\mathrm{c}_{0}}{\eta_{0}} - \frac{I^\mathrm{c}_{60}}{\eta_{60}} ) + \frac{I^\mathrm{c}_{60}}{\eta_{60}} ( \frac{I^\mathrm{c}_{60}}{\eta_{60}} -\frac{I^\mathrm{c}_{120}}{\eta_{120}} ) + \frac{I^\mathrm{c}_{120}}{\eta_{120}} ( \frac{I^\mathrm{c}_{120}}{\eta_{120}} - \frac{I^\mathrm{c}_{0}}{\eta_{0}} )}}{\frac{I^\mathrm{c}_{0}}{\eta_{0}} + \frac{I^\mathrm{c}_{60}}{\eta_{60}} + \frac{I^\mathrm{c}_{120}}{\eta_{120}}}.\label{eq:P430_2}\\
P &=& \frac{2\sqrt{I^\mathrm{c}_{0}/\eta_{0} (I^\mathrm{c}_{0}/\eta_{0} - I^\mathrm{c}_{60}/\eta_{60}) + I^\mathrm{c}_{60}/\eta_{60} (I^\mathrm{c}_{60}/\eta_{60} -I^\mathrm{c}_{120}/\eta_{120}) + I^\mathrm{c}_{120}/\eta_{120} (I^\mathrm{c}_{120}/\eta_{120} - I^\mathrm{c}_{0}/\eta_{0})}}{I^\mathrm{c}_{0}/\eta_{0} + I^\mathrm{c}_{60}/\eta_{60} + I^\mathrm{c}_{120}/\eta_{120}} \label{eq:P430_2}\\
\Psi_{P} &=& \frac{1}{2} \arctan \frac{\sqrt{3} ( I^\mathrm{c}_{60}/\eta_{60} - I^\mathrm{c}_{120}/\eta_{120} )}{2 I^\mathrm{c}_{0}/\eta_{0} - I^\mathrm{c}_{60}/\eta_{60} - I^\mathrm{c}_{120}/\eta_{120}}. \label{eq:A430_2}
\end{eqnarray}

\clearpage
\section{Additional Examples of Smear Correction}\label{sec:smear_ex}
% Figure~\ref{fig:smear}(b) and~\ref{fig:smear}(c)
% second column Figure~\ref{fig:dop}

% Fig: Smear Example 1 ---------------------------------------
\begin{figure*}[hp!]
\centering
\includegraphics[width=0.95\textwidth]{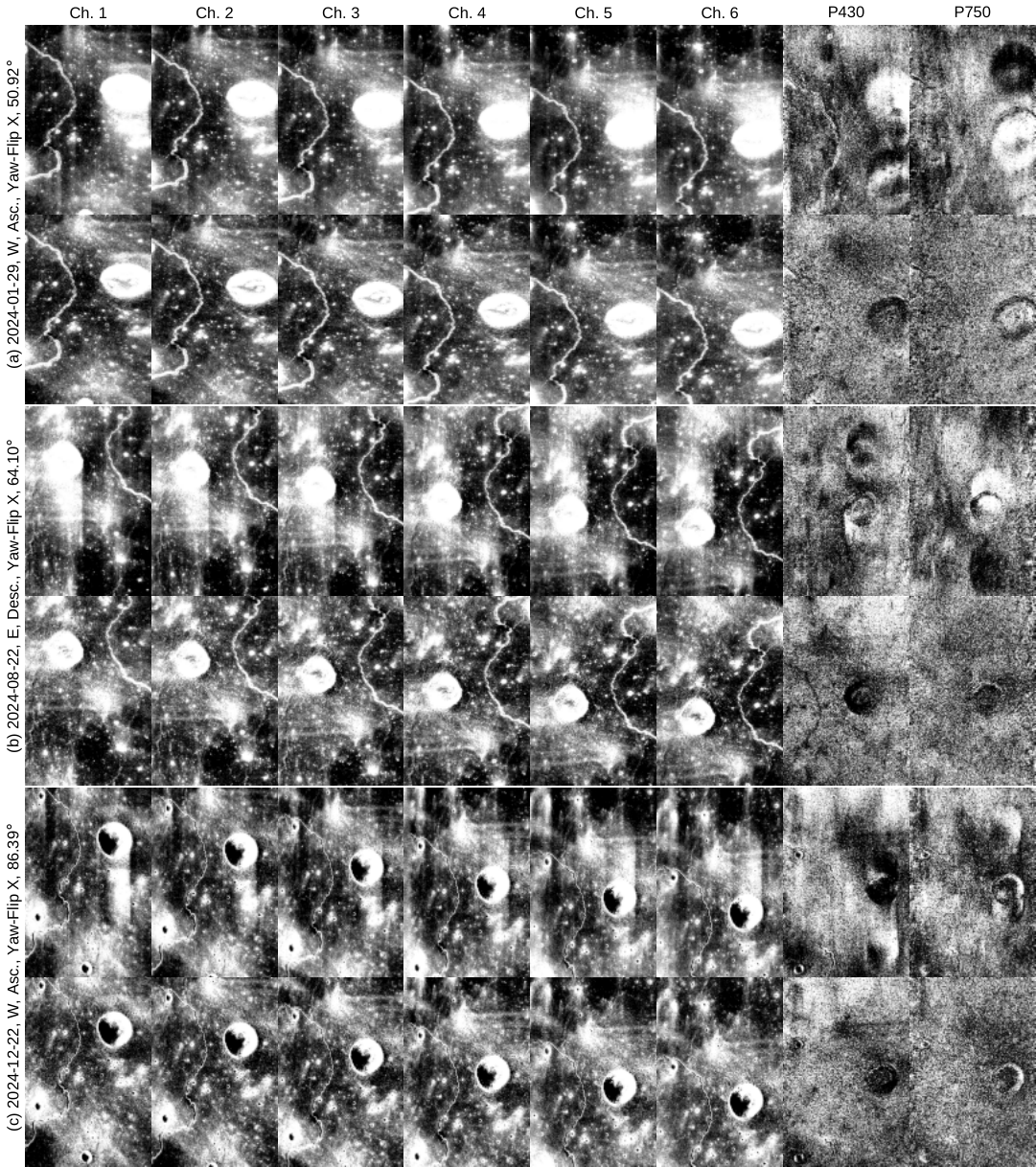}
\caption{Typical cases of effective smear mitigation for the Marius~B crater. Three examples are provided; in each panel, the top and bottom rows show the images before and after smear correction, respectively. The left six columns show display intensity images from each channel, while the 7th and 8th columns present the degree of linear polarization images at 430~nm and 750~nm, respectively. Labels indicate the observation date, camera-facing direction (East or West), orbit direction (Ascending or Descending), yaw-flip status, and average phase angle.\label{fig:smear_ex1}}
\end{figure*}
% ------------------------------------------------------------

% Fig: Smear Example 2 ---------------------------------------
\begin{figure*}[hp!]
\centering
\includegraphics[width=0.95\textwidth]{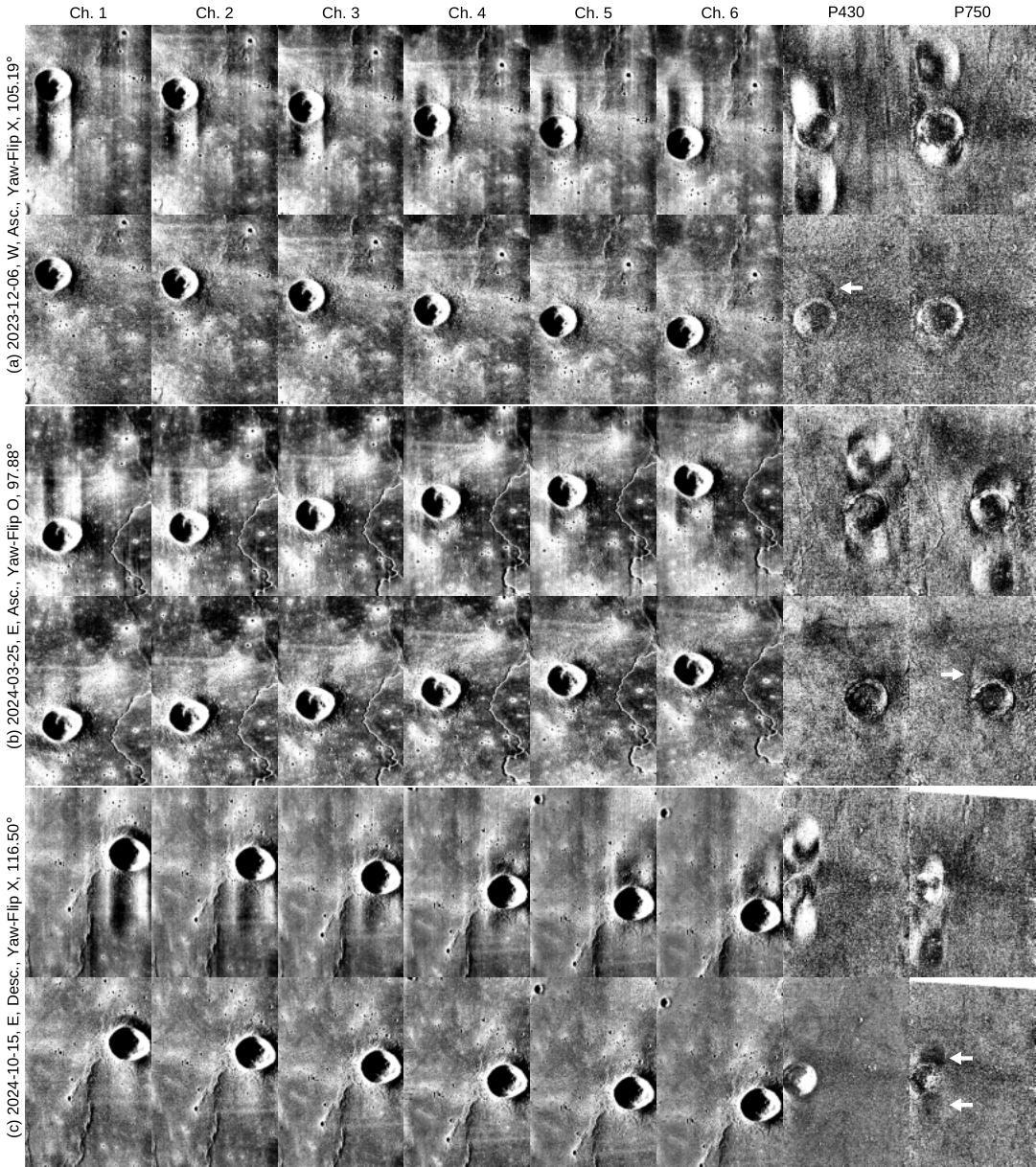}
\caption{Same as Figure~\ref{fig:smear_ex1}, but showing the poor-performing cases of smear correction for the Marius~B crater. White arrows indicate under-corrected (Panel~(a)) and over-corrected (Panels~(b) and~(c)) regions. These correction errors are not distinguishable in the intensity images but are evident in the degree of linear polarization images.\label{fig:smear_ex2}}
\end{figure*}
% ------------------------------------------------------------

\clearpage
%% For this sample we use BibTeX plus aasjournalv7.bst to generate the
\bibliography{PolCamImgCal}{}

@ARTICLE{Iglesias2015,
       author = {{Iglesias}, Francisco A. and {Feller}, Alex and {Nagaraju}, Krishnappa},
        title = "{Smear correction of highly variable, frame-transfer CCD images with application to polarimetry}",
      journal = {\ao},
         year = 2015,
        month = jul,
       volume = {54},
       number = {19},
        pages = {5970-5975},
          doi = {10.1364/AO.54.005970},
archivePrefix = {arXiv},
       eprint = {1506.03706},
 primaryClass = {astro-ph.IM},
       adsurl = {https://ui.adsabs.harvard.edu/abs/2015ApOpt..54.5970I}
}

@INPROCEEDINGS{Feller2014,
       author = {{Feller}, A. and {Iglesias}, F.~A. and {Nagaraju}, K. and {Solanki}, S.~K. and {Ihle}, S.},
        title = "{Fast Solar Polarimeter: Description and First Results}",
    booktitle = {Solar Polarization 7},
         year = 2014,
       editor = {{Nagendra}, K.~N. and {Stenflo}, J.~O. and {Qu}, Z.~Q. and {Sampoorna}, M.},
       series = {ASP Conf. Ser.},
       volume = {489},
        month = oct,
        pages = {271},
       adsurl = {https://ui.adsabs.harvard.edu/abs/2014ASPC..489..271F}
}

@ARTICLE{Iglesias2016,
       author = {{Iglesias}, F.~A. and {Feller}, A. and {Nagaraju}, K. and {Solanki}, S.~K.},
        title = "{High-resolution, high-sensitivity, ground-based solar spectropolarimetry with a new fast imaging polarimeter. I. Prototype characterization}",
      journal = {\aap},
         year = 2016,
        month = may,
       volume = {590},
          eid = {A89},
        pages = {A89},
          doi = {10.1051/0004-6361/201628376},
archivePrefix = {arXiv},
       eprint = {1604.01521},
 primaryClass = {astro-ph.IM},
       adsurl = {https://ui.adsabs.harvard.edu/abs/2016A&A...590A..89I}
}

@ARTICLE{Baek2025,
       author = {{Baek}, Kilho and {Kim}, Sungsoo S. and {Jeong}, Minsup and {Choi}, Young-Jun},
        title = "{On-Orbit Calibration of Danuri/PolCam. I. Geometric Calibration}",
      journal = {Journal of the Korean Astronomical Society},
         year = 2025,
        month = dec,
       volume = {58},
        pages = {291-303},
          doi = {10.5303/JKAS.2025.58.2.291},
archivePrefix = {arXiv},
       eprint = {2512.05330},
 primaryClass = {astro-ph.EP},
       adsurl = {https://ui.adsabs.harvard.edu/abs/2025JKAS...58..291B}
}

@ARTICLE{Jeong2023,
       author = {{Jeong}, Minsup and {Choi}, Young-Jun and {Kang}, Kyung-In and {Moon}, Bongkon and {Gu}, Bonju and {Kim}, Sungsoo S. and {Sim}, Chae Kyung and {Lee}, Dukhang and {Shkuratov}, Yuriy G. and {Videen}, Gorden and {Kaydash}, Vadym},
        title = "{Preflight Calibration Results of Wide-Angle Polarimetric Camera (PolCam) onboard Korean Lunar Orbiter, Danuri}",
      journal = {Journal of the Korean Astronomical Society},
         year = 2023,
        month = dec,
       volume = {56},
        pages = {293-299},
          doi = {10.5303/JKAS.2023.56.2.293},
       adsurl = {https://ui.adsabs.harvard.edu/abs/2023JKAS...56..293J}
}

@ARTICLE{Jeong2015,
       author = {{Jeong}, Minsup and {Kim}, Sungsoo S. and {Garrick-Bethell}, Ian and {Park}, So-Myoung and {Sim}, Chae Kyung and {Jin}, Ho and {Min}, Kyoung Wook and {Choi}, Young-Jun},
        title = "{Multi-band Polarimetry of the Lunar Surface. I. Global Properties}",
      journal = {\apjs},
         year = 2015,
        month = nov,
       volume = {221},
       number = {1},
          eid = {16},
        pages = {16},
          doi = {10.1088/0067-0049/221/1/16},
       adsurl = {https://ui.adsabs.harvard.edu/abs/2015ApJS..221...16J}
}

@ARTICLE{Sim2020,
       author = {{Sim}, Chae Kyung and {Kim}, Sungsoo S. and {Jeong}, Minsup and {Choi}, Young-Jun and {Shkuratov}, Yuriy G.},
        title = "{Observational Strategy for KPLO/PolCam Measurements of the Lunar Surface from Orbit}",
      journal = {\pasp},
         year = 2020,
        month = jan,
       volume = {132},
       number = {1007},
          eid = {015004},
        pages = {015004},
          doi = {10.1088/1538-3873/ab523d},
       adsurl = {https://ui.adsabs.harvard.edu/abs/2020PASP..132a5004S}
}

@ARTICLE{Jeon2024,
       author = {{Jeon}, Moon-Jin and {Cho}, Young-Ho and {Kim}, Eunhyeuk and {Kim}, Dong-Gyu and {Song}, Young-Joo and {Hong}, SeungBum and {Bae}, Jonghee and {Bang}, Jun and {Yim}, Jo Ryeong and {Kim}, Dae-Kwan},
        title = "{Korea Pathfinder Lunar Orbiter (KPLO) Operation: From Design to Initial Results}",
      journal = {Journal of Astronomy and Space Sciences},
         year = 2024,
        month = mar,
       volume = {41},
       number = {1},
        pages = {43-60},
          doi = {10.5140/JASS.2024.41.1.43},
       adsurl = {https://ui.adsabs.harvard.edu/abs/2024JASS...41...43J}
}

@ARTICLE{Song2023,
       author = {{Song}, Young-Joo and {Bang}, Jun and {Bae}, Jonghee and {Hong}, SeungBum},
        title = "{Lunar Orbit acquisition of the Korea Pathfinder lunar orbiter: Design reference vs actual flight results}",
      journal = {Acta Astronautica},
         year = 2023,
        month = dec,
       volume = {213},
        pages = {336-343},
          doi = {10.1016/j.actaastro.2023.09.021},
       adsurl = {https://ui.adsabs.harvard.edu/abs/2023AcAau.213..336S}
}

@ARTICLE{Song2026,
       author = {{Song}, Young-Joo and {Hong}, SeungBum and {Bang}, Jun and {Bae}, Jonghee and {Jeon}, Moon-Jin and {Chung}, Soyoung and {Fuller}, Shane and {Stuit}, Timothy},
        title = "{KPLO's conjunction mitigation in lunar orbit: Operational results and strategic insights from international collaboration}",
      journal = {Acta Astronautica},
         year = 2026,
        month = apr,
       volume = {241},
        pages = {637-653},
          doi = {10.1016/j.actaastro.2025.11.061},
       adsurl = {https://ui.adsabs.harvard.edu/abs/2026AcAau.241..637S}
}

@INCOLLECTION{Shkuratov2025,
    author = {{Shkuratov}, Yuriy and {Videen}, Gorden and {Kaydash}, Vadym},
    title = {Lunar polarimetry: Observations with telescopes as well as laboratory, computer, and theoretical modelings},
    booktitle = {Optics of the Moon},
    publisher = {Elsevier},
    pages = {595-728},
    year = {2025},
    isbn = {978-0-12-817972-7},
    doi = {10.1016/B978-0-12-817972-7.00006-3},
    url = {https://www.sciencedirect.com/science/article/pii/B9780128179727000063}
}

@ARTICLE{Shkuratov2026a,
       author = {{Shkuratov}, Yuriy and {Kaydash}, Vadym and {Head}, James and {Baek}, Kilho and {Kim}, Sungsoo S. and {Opanasenko}, Nickolay and {Surkov}, Yehor and {Jeong}, Minsup and {Choi}, Young-Jun and {Sim}, Chae Kyung and {Lee}, Dukhang and {Kreslavsky}, Michael and {Farrand}, William H. and {Fassett}, Caleb and {Videen}, Gorden},
        title = "{Characterization of the surface of the Ina irregular Mare patch: A synthesis of lunar optical data, including Danuri polarimetric measurements}",
      journal = {\icarus},
         year = 2026,
        month = jan,
       volume = {444},
          eid = {116796},
        pages = {116796},
          doi = {10.1016/j.icarus.2025.116796},
       adsurl = {https://ui.adsabs.harvard.edu/abs/2026Icar..44416796S}
}

@ARTICLE{Shkuratov2026b,
       author = {{Shkuratov}, Yuriy and {Kaydash}, Vadym and {Head}, James and {Baek}, Kilho and {Kim}, Sungsoo S. and {Qiao}, Le and {Velichko}, Sergey and {Surkov}, Yehor and {Opanasenko}, Nickolay and {Jeong}, Minsup and {Choi}, Young-Jun and {Sim}, Chae Kyung and {Kreslavsky}, Michael and {Farrand}, William and {Videen}, Gorden},
        title = "{The Maskelyne irregular mare patch as interpreted with lunar optical data, including Danuri polarimetric measurements}",
      journal = {\icarus},
         year = 2026,
        month = jul,
       volume = {453},
          eid = {117059},
        pages = {117059},
          doi = {10.1016/j.icarus.2026.117059},
       adsurl = {https://ui.adsabs.harvard.edu/abs/2026Icar..45317059S}
}

@ARTICLE{Shkuratov2026c,
       author = {{Shkuratov}, Yuriy and {Kaydash}, Vadym and {Head}, James and {Baek}, Kilho and {Kim}, Sungsoo S. and {Velichko}, Sergey and {Surkov}, Yehor and {Jeong}, Minsup and {Choi}, Young-Jun and {Sim}, Chae Kyung and et al.},
        title = "{Lunar Sosigenes IMP as analyzed with orbital spectrophotometry and polarimetry}",
      journal = {\icarus},
         year = 2026,
        month = dec,
       volume = {460},
          eid = {117285},
        pages = {117285},
          doi = {10.1016/j.icarus.2026.117285},
       adsurl = {https://ui.adsabs.harvard.edu/abs/2026Icar..46017285S}
}

@ARTICLE{Umov1905,
    author = {{Umov}, Nikolay Alekseevich},
    title = {Chromatische depolarisation durch lichtzerstreung},
    journal = {Phis. Zeits},
    year = 1905,
    volume = {6},
    pages = {674-676}
}

@ARTICLE{Dollfus1971,
       author = {{Dollfus}, A. and {Bowell}, E. and {Titulaer}, C.},
        title = "{Polarimetric Properties of the Lunar Surface and its Interpre- tation. Part II. Terrestrial Samples in Orange Light}",
      journal = {\aap},
         year = 1971,
        month = feb,
       volume = {10},
        pages = {450},
       adsurl = {https://ui.adsabs.harvard.edu/abs/1971A&A....10..450D}
}

@ARTICLE{Wolff1980,
       author = {{Wolff}, M.},
        title = "{Theory and application of the polarization-albedo rules}",
      journal = {\icarus},
         year = 1980,
        month = dec,
       volume = {44},
       number = {3},
        pages = {780-792},
          doi = {10.1016/0019-1035(80)90144-X},
       adsurl = {https://ui.adsabs.harvard.edu/abs/1980Icar...44..780W}
}

@ARTICLE{Shkuratov1992,
       author = {{Shkuratov}, Iu. G. and {Opanasenko}, N.~V.},
        title = "{Polarimetric and photometric properties of the moon: Telescope observation and laboratory simulation  2. The positive polarization}",
      journal = {\icarus},
         year = 1992,
        month = oct,
       volume = {99},
       number = {2},
        pages = {468-484},
          doi = {10.1016/0019-1035(92)90161-Y},
       adsurl = {https://ui.adsabs.harvard.edu/abs/1992Icar...99..468S}
}

@ARTICLE{Lipskii1967,
       author = {{Lipskii}, Yu. N. and {Pospergelis}, M.~M.},
        title = "{Some Results of Measurements of the Total Stokes Vector for Details of the Lunar Surface.}",
      journal = {\azh},
         year = 1967,
        month = jan,
       volume = {44},
        pages = {410},
       adsurl = {https://ui.adsabs.harvard.edu/abs/1967AZh....44..410L}
}

@ARTICLE{Shkuratov2011,
       author = {{Shkuratov}, Y. and {Kaydash}, V. and {Korokhin}, V. and {Velikodsky}, Y. and {Opanasenko}, N. and {Videen}, G.},
        title = "{Optical measurements of the Moon as a tool to study its surface}",
      journal = {\planss},
         year = 2011,
        month = oct,
       volume = {59},
       number = {13},
        pages = {1326-1371},
          doi = {10.1016/j.pss.2011.06.011},
       adsurl = {https://ui.adsabs.harvard.edu/abs/2011P&SS...59.1326S}
}

@ARTICLE{Feng2013,
       author = {{Feng}, Weiwei and {Chen}, Ligang},
        title = "{Smear correction of frame transfer CCD camera based on the integrating sphere}",
      journal = {Optik},
         year = 2013,
        month = jul,
       volume = {124},
       number = {14},
        pages = {1805-1807},
          doi = {10.1016/j.ijleo.2012.05.043},
       adsurl = {https://ui.adsabs.harvard.edu/abs/2013Optik.124.1805F}
}

@INPROCEEDINGS{Knox2007,
       author = {{Knox}, K.},
        title = "{Recovering Saturated Pixels Blurred by CCD Image Smear}",
    booktitle = {Advanced Maui Optical and Space Surveillance Technologies Conference},
         year = 2007,
       editor = {{Ryan}, S.},
        month = jan,
          eid = {E59},
        pages = {E59},
       adsurl = {https://ui.adsabs.harvard.edu/abs/2007amos.confE..59K}
}

@ARTICLE{Ruyten1999,
       author = {{Ruyten}, Wim},
        title = "{Smear correction forframe transfer charge-coupled-device cameras}",
      journal = {Optics Letters},
         year = 1999,
        month = jul,
       volume = {24},
       number = {13},
        pages = {878-880},
          doi = {10.1364/OL.24.000878},
       adsurl = {https://ui.adsabs.harvard.edu/abs/1999OptL...24..878R}
}

@ARTICLE{Powell1999,
       author = {{Powell}, Keith and {Chana}, Deeph and {Fish}, David and {Thompson}, Chris},
        title = "{Restoration and Frequency Analysis of Smeared CCD Images}",
      journal = {\ao},
         year = 1999,
        month = mar,
       volume = {38},
       number = {8},
        pages = {1343-1347},
          doi = {10.1364/AO.38.001343},
       adsurl = {https://ui.adsabs.harvard.edu/abs/1999ApOpt..38.1343P}
}

@ARTICLE{Denevi2018,
       author = {{Denevi}, Brett W. and {Chabot}, Nancy L. and {Murchie}, Scott L. and {Becker}, Kris J. and {Blewett}, David T. and {Domingue}, Deborah L. and {Ernst}, Carolyn M. and {Hash}, Christopher D. and {Hawkins}, S. Edward and {Keller}, Mary R. and {Laslo}, Nori R. and {Nair}, Hari and {Robinson}, Mark S. and {Seelos}, Frank P. and {Stephens}, Grant K. and {Turner}, F. Scott and {Solomon}, Sean C.},
        title = "{Calibration, Projection, and Final Image Products of MESSENGER's Mercury Dual Imaging System}",
      journal = {\ssr},
         year = 2018,
        month = feb,
       volume = {214},
       number = {1},
          eid = {2},
        pages = {2},
          doi = {10.1007/s11214-017-0440-y},
       adsurl = {https://ui.adsabs.harvard.edu/abs/2018SSRv..214....2D}
}

@ARTICLE{Hawkins2007,
       author = {{Hawkins}, S. Edward and {Boldt}, John D. and {Darlington}, Edward H. and {Espiritu}, Raymond and {Gold}, Robert E. and {Gotwols}, Bruce and {Grey}, Matthew P. and {Hash}, Christopher D. and {Hayes}, John R. and {Jaskulek}, Steven E. and {Kardian}, Charles J. and {Keller}, Mary R. and {Malaret}, Erick R. and {Murchie}, Scott L. and {Murphy}, Patricia K. and {Peacock}, Keith and {Prockter}, Louise M. and {Reiter}, R. Alan and {Robinson}, Mark S. and {Schaefer}, Edward D. and {Shelton}, Richard G. and {Sterner}, Raymond E. and {Taylor}, Howard W. and {Watters}, Thomas R. and {Williams}, Bruce D.},
        title = "{The Mercury Dual Imaging System on the MESSENGER Spacecraft}",
      journal = {\ssr},
         year = 2007,
        month = aug,
       volume = {131},
       number = {1-4},
        pages = {247-338},
          doi = {10.1007/s11214-007-9266-3},
       adsurl = {https://ui.adsabs.harvard.edu/abs/2007SSRv..131..247H}
}

@ARTICLE{Murchie1999,
       author = {{Murchie}, Scott and {Robinson}, Mark and {Hawkins}, S. Edward and {Harch}, Ann and {Helfenstein}, Paul and {Thomas}, Peter and {Peacock}, Keith and {Owen}, William and {Heyler}, Gene and {Murphy}, Patricia and {Darlington}, E.~H. and {Keeney}, Allen and {Gold}, Robert and {Clark}, Beth and {Izenberg}, Noam and {Bell}, James F. and {Merline}, William and {Veverka}, Joseph},
        title = "{Inflight Calibration of the NEAR Multispectral Imager}",
      journal = {\icarus},
         year = 1999,
        month = jul,
       volume = {140},
       number = {1},
        pages = {66-91},
          doi = {10.1006/icar.1999.6118},
       adsurl = {https://ui.adsabs.harvard.edu/abs/1999Icar..140...66M}
}

@ARTICLE{Kodama2010,
       author = {{Kodama}, Shinsuke and {Ohtake}, Makiko and {Yokota}, Yasuhiro and {Iwasaki}, Akira and {Haruyama}, Junichi and {Matsunaga}, Tsuneo and {Nakamura}, Ryosuke and {Demura}, Hirohide and {Hirata}, Naru and {Sugihara}, Takamitsu and {Yamamoto}, Yasuji},
        title = "{Characterization of Multiband Imager Aboard SELENE. Pre-flight and In-flight Radiometric Calibration}",
      journal = {\ssr},
         year = 2010,
        month = jul,
       volume = {154},
       number = {1-4},
        pages = {79-102},
          doi = {10.1007/s11214-010-9661-z},
       adsurl = {https://ui.adsabs.harvard.edu/abs/2010SSRv..154...79K}
}

@ARTICLE{Ishiguro2010,
       author = {{Ishiguro}, Masateru and {Nakamura}, Ryosuke and {Tholen}, David J. and {Hirata}, Naru and {Demura}, Hirohide and {Nemoto}, Etsuko and {Nakamura}, Akiko M. and {Higuchi}, Yuta and {Sogame}, Akito and {Yamamoto}, Aya and {Kitazato}, Kohei and {Yokota}, Yasuhiro and {Kubota}, Takashi and {Hashimoto}, Tatsuaki and {Saito}, Jun},
        title = "{The Hayabusa Spacecraft Asteroid Multi-band Imaging Camera (AMICA)}",
      journal = {\icarus},
         year = 2010,
        month = jun,
       volume = {207},
       number = {2},
        pages = {714-731},
          doi = {10.1016/j.icarus.2009.12.035},
archivePrefix = {arXiv},
       eprint = {0912.4797},
 primaryClass = {astro-ph.EP},
       adsurl = {https://ui.adsabs.harvard.edu/abs/2010Icar..207..714I}
}

@ARTICLE{Hoyer2020,
       author = {{Hoyer}, S. and {Guterman}, P. and {Demangeon}, O. and {Sousa}, S.~G. and {Deleuil}, M. and {Meunier}, J.~C. and {Benz}, W.},
        title = "{Expected performances of the Characterising Exoplanet Satellite (CHEOPS). III. Data reduction pipeline: architecture and simulated performances}",
      journal = {\aap},
         year = 2020,
        month = mar,
       volume = {635},
          eid = {A24},
        pages = {A24},
          doi = {10.1051/0004-6361/201936325},
archivePrefix = {arXiv},
       eprint = {1909.08363},
 primaryClass = {astro-ph.IM},
       adsurl = {https://ui.adsabs.harvard.edu/abs/2020A&A...635A..24H}
}

@ARTICLE{Korokhin2005,
       author = {{Korokhin}, V.~V. and {Velikodsky}, Yu. I.},
        title = "{Parameters of the positive polarization maximum of the Moon: Mapping}",
      journal = {Solar System Research},
         year = 2005,
        month = jan,
       volume = {39},
       number = {1},
        pages = {45-53},
          doi = {10.1007/s11208-005-0016-3},
       adsurl = {https://ui.adsabs.harvard.edu/abs/2005SoSyR..39...45K}
}

@ARTICLE{Shkuratov2007,
       author = {{Shkuratov}, Yuriy and {Opanasenko}, Nikolay and {Zubko}, Evgenij and {Grynko}, Yevgen and {Korokhin}, Viktor and {Pieters}, Carl{\'e} and {Videen}, Gorden and {Mall}, Urs and {Opanasenko}, Alexander},
        title = "{Multispectral polarimetry as a tool to investigate texture and chemistry of lunar regolith particles}",
      journal = {\icarus},
         year = 2007,
        month = apr,
       volume = {187},
       number = {2},
        pages = {406-416},
          doi = {10.1016/j.icarus.2006.10.012},
       adsurl = {https://ui.adsabs.harvard.edu/abs/2007Icar..187..406S}
}

@ARTICLE{Wang2024,
       author = {{Wang}, Weinan and {Ping}, Jinsong and {Zhang}, Wenzhao and {Wang}, Mingyuan and {Ye}, Hanlin and {Han}, Xingwei and {Kou}, Songfeng},
        title = "{A New Method for Ground-Based Optical Polarization Observation of the Moon}",
      journal = {Sensors},
         year = 2024,
        month = apr,
       volume = {24},
       number = {8},
          eid = {2580},
        pages = {2580},
          doi = {10.3390/s24082580},
       adsurl = {https://ui.adsabs.harvard.edu/abs/2024Senso..24.2580W}
}

@INPROCEEDINGS{Nelson2014,
       author = {{Nelson}, D.~M. and {Koeber}, S.~D. and {Daud}, K. and {Robinson}, M.~S. and {Watters}, T.~R. and {Banks}, M.~E. and {Williams}, N.~R.},
        title = "{Mapping Lunar Maria Extents and Lobate Scarps Using LROC Image Products}",
    booktitle = {45th Annual Lunar and Planetary Science Conference},
         year = 2014,
       series = {Lunar and Planetary Science Conference},
        month = mar,
        pages = {2861},
       adsurl = {https://ui.adsabs.harvard.edu/abs/2014LPI....45.2861N}
}

@BOOK{Hapke2012,
       author = {{Hapke}, Bruce},
        title = "{Theory of Reflectance and Emittance Spectroscopy}",
         year = 2012,
    publisher = {Cambridge University Press},
          doi = {10.1017/CBO9781139025683},
       adsurl = {https://ui.adsabs.harvard.edu/abs/2012tres.book.....H}
}

@ARTICLE{Shkuratov2012,
       author = {{Shkuratov}, Y. and {Kaydash}, V. and {Korokhin}, V. and {Velikodsky}, Y. and {Petrov}, D. and {Zubko}, E. and {Stankevich}, D. and {Videen}, G.},
        title = "{A critical assessment of the Hapke photometric model}",
      journal = {\jqsrt},
         year = 2012,
        month = dec,
       volume = {113},
       number = {18},
        pages = {2431-2456},
          doi = {10.1016/j.jqsrt.2012.04.010},
       adsurl = {https://ui.adsabs.harvard.edu/abs/2012JQSRT.113.2431S}
}
\bibliographystyle{aasjournalv7}


\end{document}